\documentclass[12pt]{article}
\usepackage{newtxtext,newtxmath}
\usepackage{graphicx}
\usepackage{hyperref}
\usepackage[letterpaper,margin=1in]{geometry}
\renewenvironment{abstract}
	{\quotation}
	{\endquotation}

\date{}

\makeatletter
\renewcommand{\fnum@figure}{\textbf{Figure \thefigure}}
\renewcommand{\fnum@table}{\textbf{Table \thetable}}
\makeatother
\usepackage{scicite}
\usepackage{url}

\def\scititle{
	On-chip multi-timescale spatiotemporal optical synchronization
}
\title{\bfseries \boldmath \scititle}

\author{
        Lida~Xu$^{1\dagger}$, Mahmoud Jalali Mehrabad$^{1\ast\dagger}$, Christopher J. Flower$^{1\dagger}$\and
        Gregory Moille$^{2}$, Alessandro Restelli$^{2}$, Daniel G. Suarez-Forero$^{1}$,  
        Yanne Chembo$^{3}$\and Sunil Mittal$^{4}$, Kartik Srinivasan$^{2}$, Mohammad Hafezi$^{1\ast}$
        \and \small$^{1}$ Joint Quantum     Institute, Quantum Technology Center, University of Maryland,College Park, MD 20742, USA
        \and \small$^{2}$ Joint Quantum Institute, University of Maryland \and \small National Institute of Standards and Technology, College Park, MD 20742, USA
        \and \small$^{3}$ Department of Electrical and Computer Engineering, Institute for Research in Electronics and Applied Physics \and \small Joint Quantum Institute, University of Maryland, College Park, MD 20742, USA 
        \and \small$^{4}$ Department of Electrical and Computer Engineering, Northeastern University, Boston, MA, USA
	\and \small$^\ast$Corresponding author. Email:  mjalalim@umd.edu, hafezi@umd.edu\and
	\small$^\dagger$These authors contributed equally to this work.
}

\begin{document} 

\maketitle

\begin{abstract} \bfseries \boldmath
Mode-locking is foundational to nonlinear optics, enabling advances in metrology, spectroscopy, and communications. However, it remains unexplored in non-harmonic, multi-timescale regimes. Here, we realize on-chip multi-timescale synchronization using topological photonics. We design a 2D lattice of 261 coupled silicon nitride ring resonators that supports nested mode-locked states with fast ($\approx$ 1 THz) single-ring and slow ($\approx$ 3 GHz) topological super-ring timescales. We observe clear signatures of multi-timescale mode-locking, including a quadratic distribution of pump noise across both azimuthal mode families, consistent with theory. These findings are supported by near-transform-limited repetition beats and the emergence of periodic temporal patterns on the slow timescale. The edge-confined states show distinct dynamics from bulk and single-ring modes, enabling clear identification. Our results establish topological frequency combs as a robust platform for independently tunable, lattice-scale synchronization, opening new directions for exploring the interplay of nonlinearity and topology in integrated photonics.

\end{abstract}
\subsection*{Short title} 
On-chip two-timescale optical
synchronization


\subsection*{Teaser} 
Simultaneous mode locking on two independently designable timescales is observed in a topological photonic lattice.

\noindent

\section*{Introduction}

In the linear regime, a propagating beam of light broadens in space and disperses in time, due to diffraction and chromatic dispersion, respectively. However, nonlinearities may create opposite dispersive or diffractive behavior, for which when they exactly counterbalance allows for mode-locking~\cite{boyd2008nonlinear}. The intensity-dependent index of refraction in nonlinear media is an early example that enables self-focusing mechanisms to counteract the diffraction-led divergence, resulting in spatially mode-locked optical solitonic solutions~\cite{segev1998self, lugiato1987spatial}. More recently, pivotal breakthroughs were reported by engineering a delicate balance between the nonlinear Kerr effect and dispersion leading to the discovery of mode-locked temporal dissipative Kerr solitons (DKS)~\cite{leo2010temporal,herr2014temporal} following the same dynamics as their spatial counterpart~\cite{chembo2013spatiotemporal}. DKS have enabled many applications~\cite{diddams2020optical} in spectroscopy~\cite{dutt2018chip}, frequency synthesis~\cite{spencer2018optical}, ranging~\cite{riemensberger2020massively}, and optical clockwork~\cite{newman2019architecture,moille2023kerr}. 

At its core, the mode-locking of a DKS can be understood as a synchronization phenomenon~\cite{wen2016self}. Recent years have witnessed substantial efforts to develop other DKS synchronization schemes such as remote resonator~\cite{jang2018synchronization,zhao2024all}, counter-propagative~\cite{yang2017counter}, pulse-~\cite{obrzud2017temporal} or continuous-wave driven~\cite{moille2023kerr} DKS, among others. However, all these synchronization schemes rely on the single, or harmonic, frequency grid defined by the DKS repetition rate. The exploration of independently designable multi-timescales and multi-spatial modes for which phase-locking may happen has remained scarce~\cite{aadhi2024mode,schwartz2013laser,wright2022nonlinear}.  
This limitation is partly due to single-resonator systems since they require harmonic synchronization even when dealing with different timescales, such as DKS-breather synchronization~\cite{cole2019subharmonic}. 

A potential solution is to create nested resonant structures with uncoupled different timescales, allowing synchronization and mode-locking to coexist in a multi-dimensional space. Long 1D chains of microring resonators were theoretically studied~\cite{marti2021slow, tusnin2023nonlinear} as a potential candidate, however, their experimental use is severely limited due to resonant frequency inhomogeneity~\cite{mittal2014topologically}.

Recently, nested-DKS in large 2D arrays, which supports topologically protected edge-mode forming a low-loss super-ring, was proposed~\cite{mittal2021topological}, followed by the experimental observation of modulation-instability combs in such 2D lattices~\cite{flower2024observation,mehrabad2025multi}. Thus, optical synchronization and multi-timescale mode-locked states should
theoretically exist but have yet to be experimentally demonstrated.

Here, we overcome these challenges using topological photonics and report the first demonstration of multi-timescale spatiotemporal mode-locking in a 2D array of more than one hundred coupled silicon nitride (SiN) ring resonators. Specifically, using the Kerr effect, we generate optical frequency combs at the edge of the topological lattice and present several direct experimental signatures of their simultaneous mode-locking in the fast ($\approx$ 1 THz) single-ring and slow ($\approx$ 3 GHz) topological super-ring timescales. We directly demonstrate the quadratic dependence of the pump noise in both single-ring and topological super-ring timescales, as expected by mode-locking theory~\cite{lei2022optical,telle2002kerr}. Our observations are further substantiated with clear signatures on the slow-time scale with near-transform-limit repetition beats on RF spectrum and the formation of periodic temporal patterns.
Moreover, we demonstrate that these mode-locking signatures are unique to combs generated in the topological edge bands, and they are in sharp contrast to the bulk counterparts, for which we establish a comprehensive pathway for their identification.

\subsection*{Concept}
\subsubsection*{The topological lattice}
Figure~\ref{Fig:intro} illustrates our system. 
The topological device consists of a 2D array of coupled ring resonators and is designed to simulate the anomalous quantum Hall (AQH) model for photons~\cite{leykam2018reconfigurable,mittal2019photonic,mittal2021topological}. The design, linear dynamics, Hamiltonian description, and band structure of the system are presented in detail in the supplementary information (SI) sections S1-S4. Here we only highlight the key characteristics of our device that are central to understanding the multi-timescale topological mode-locking mechanism. 

One of the key characteristics of such a photonic AQH lattice is the existence of an edge band, spectrally located between two bulk bands, near each of the longitudinal mode resonances of the individual rings $\mu$, as shown in Figure~\ref{Fig:intro}B. This repeating bulk-edge-bulk spectrum is numerically calculated by simulating weak broadband light coupled into the device’s input port while monitoring the drop port. Another key feature of the system is that in sharp contrast to the bulk modes, the edge mode resonances within the edge band are spatially confined to the boundary of the lattice with a well-defined round trip time, as shown in Figure~\ref{Fig:intro}C. The third important feature is the existence of two timescales in the system, which arise from the fast $\tau_{\rm{F}}$ and slow $\tau_{\rm{S}}$ round trip times associated with the individual ring and the super ring, respectively.

\subsubsection*{Formation and detection of topological mode-locked states}

\begin{figure*}[t]
    \centering
    \includegraphics[width=0.99\textwidth]{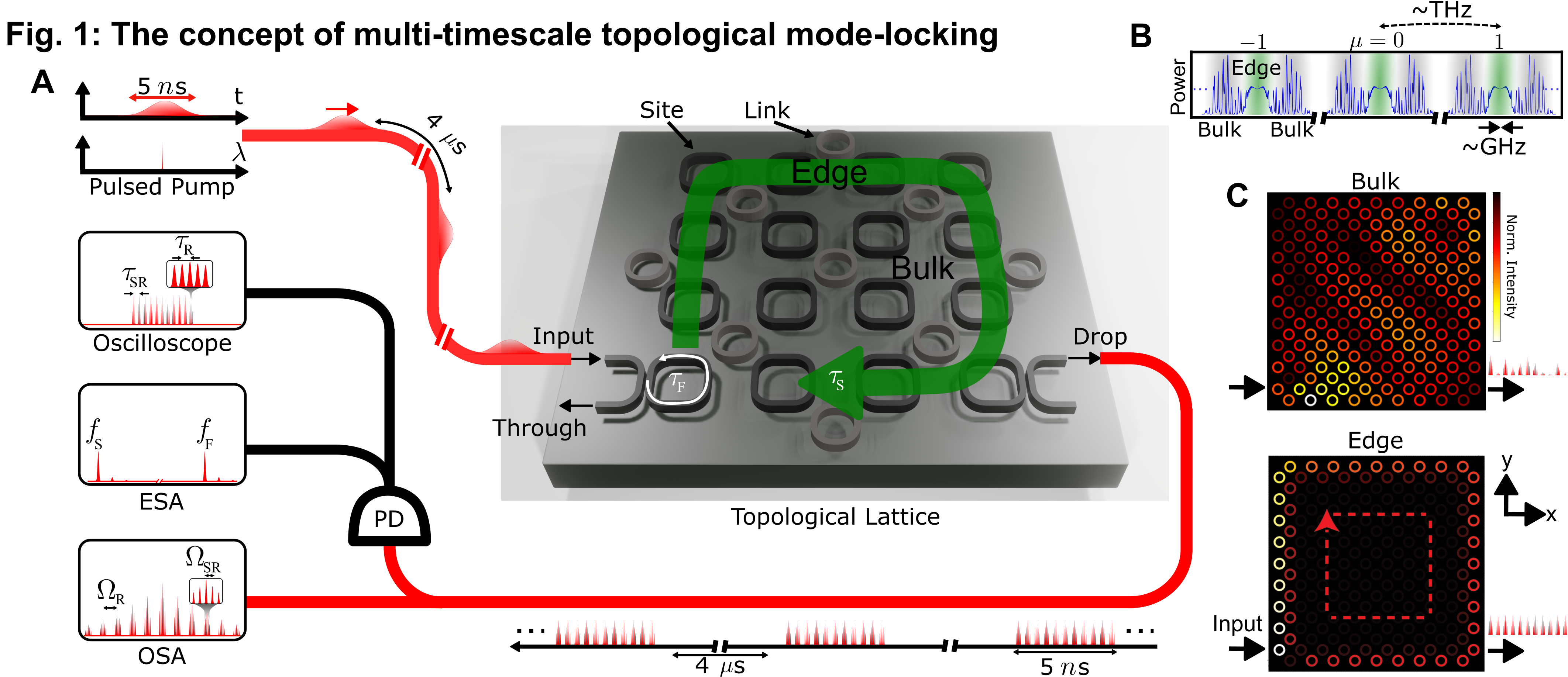}
    \caption{\textbf{The concept of topological multi timescale mode-locking. }\textbf{A,} Schematic of the experimental setup for the generation and detection of topological temporal mode-locked states in a simplified 2D array of coupled Kerr-ring resonators (see Figure~\ref{Fig:band_Linear} for the actual device). A tunable pulsed pump with a 5 ns pulse duration and 4 $\mu$s repetition period is coupled into the lattice at the input port and circulates around the edge of the 2D AQH SiN lattice (here only the clock-wise mode is shown). The generated frequency combs are directly analyzed using an optical spectrum analyzer (OSA), revealing the comb spacing of the single rings $\nu_{\rm{F}}$ and the nested comb spacing of the super ring, $\nu_{\rm{S}}$. The frequency combs are also recorded with a photodiode (PD) and analyzed with an oscilloscope and electrical spectrum analyzer (ESA), respectively, revealing the repetition rate of the combs and their time domain dynamics. \textbf{B,} 
    Simulated linear transmission of the device. The edge and bulk bands of the spectrum are shaded by green and gray, respectively. \textbf{C,}  Simulated spatial profile of representative edge (bottom) and (top) bulk modes in the linear regime (shown only for the clockwise excitation.)  
    }
    \label{Fig:intro}
\end{figure*}

Figure~\ref{Fig:intro}A illustrates our scheme for the generation of nested topologically mode-locked states of light. The starting point is sending a pulsed pump with a 5 ns pulse duration and 4 $\mu$s period (250 kHz repetition rate), in resonance with the edge band of the lattice, into the input port of the device. The Kerr nonlinearity of SiN leads to the formation of frequency combs through a stimulated four-wave mixing process~\cite{flower2024observation}. We note that using a 5 ns pulsed pump provides the peak power necessary for reaching the high comb generation threshold in such a large-scale resonant structure while maintaining a low average power to avoid unwanted thermal effects. Importantly, 5 ns is at least an order of magnitude longer than the slowest time scales in our device, that is the $\tau_{\rm{S}} \approx$ 300~ps, as well as the transient time $\approx$ 400~ps, constituting a quasi-continuous wave excitation regime for our experiment. The discussion of transient time can be found in SI section S19.

Similar to single-ring counterparts~\cite{herr2014temporal}, when the nonlinearity and the dispersion in the system are delicately balanced, the topological optical frequency combs can become mode-locked~\cite{mittal2021topological}. In this scenario, the output of the lattice is pulses of light with periodic patterns. We note that in a more general setting, such periodic patterns don't have to be hyperbolic secant, that is the solution of solitons. Importantly, in contrast to the conventional single-ring microcombs, the periodicity of the output pulses has two timescales, one originating from mode-locking in the fast timescale $\tau_{\rm{F}}$, and one from the slow timescale $\tau_{\rm{S}}$. We note that, importantly, these two timescales can be designed independently (see the SI section S2 for details). Signatures of such two-timescale periodicity can be detected directly by analyzing the output port of a photodiode as shown in Figure~\ref{Fig:intro}A.

For detection, the generated frequency combs are analyzed using an OSA, revealing the comb spacing of the single rings $\nu_{\rm F}=1/\tau_{\rm F}$ and the nested comb spacing of the super ring $\nu_{\rm S}=1/\tau_{\rm S}$. We index sing-ring modes with $\mu$ and super-ring modes with $\sigma$. The combs are also recorded with a PD and analyzed with an oscilloscope and electrical spectrum analyzer (ESA), respectively, revealing the repetition rate of the combs and their time domain dynamics.

Moreover, in contrast to conventional single rings, our topological lattice gives rise to a band structure with distinct edge and bulk bands. For our AQH lattice, as shown in Figure~\ref{Fig:intro}B, this is in the form of an edge band (green) with two bulk bands (gray) on its side, with the width of the bands and the number of edge modes set by the lattice parameters (see the SI sections S1-S3 for details). Importantly, only the edge band has a well-defined round-trip time, and the bulk bands lack such character. As such, temporal mode-locking can only be expected for the edge combs. 
Throughout this work, we investigate the characteristics of mode-locking in the edge part of the spectrum and show how they substantially differ compared to their bulk counterparts. Details of the complications, limitations and potential future improvements of our work is presented in the SI section S22.

\section*{Results}
\subsection*{Optical Signatures of Edge and Bulk frequency combs}
We begin by classifying optical characteristics such as combs' spectral envelopes, spatial profiles, nested structures, and optical linewidths which are crucial in the identification of bulk and edge combs in our system. 

\subsubsection*{Optical spectrum profiles}
 
We generate topological frequency combs by pumping the array above the optical parametric oscillation (OPO) threshold using 185 mW of average power (corresponding to on-chip peak power $\approx$ 148 W). The details of the nonlinear measurement setup, the pump spectrum, and the OPO threshold characterization are presented in the SI sections S5, S10, and S13.

To classify the differences between edge and bulk combs, we focus on a longitudinal mode $\mu = 0$ at 1547.8 nm. We note that coupling the light into the input port selectively only excites the counter-clockwise (CCW) edge modes, which is the edge mode we focus on from here on. We choose the wavelength range of the pump, starting from the bulk at the shorter-wavelength part of the spectrum (1547.6 nm) up to the longer-wavelength part (1548.1 nm), to cover the entire edge band as well as parts of the adjacent two bulk bands. The complete experimental linear characterization of the device is presented in the SI section S4. First, the generated combs are sent to a low-resolution (40 pm), broadband (1300 nm to 1800 nm) OSA. The results are shown in Figure~\ref{Fig:osa}A. We observe that in contrast to the bulk which does not have any appreciable envelope pattern, pumping the edge band of the lattice generates a comb with a triangular-like envelope in wavelength. The spacing between the adjacent comb teeth is about 6.3 nm, which is the single-ring free spectral range $\nu_{\rm{F}}$. One observation is that the bulk comb is broader than edge combs, while the variation of the broadness of the edge combs is related to the linear transmission power at different parts of the edge band, as shown in detail in the SI sections S4 and S13. Apart from broadness, we show that the threshold of bulk combs is higher than that of edge combs, as shown in SI section S17. The transmission of edge combs is also typically higher than bulk, as shown in SI section S13.

\subsubsection*{Spatial Profiles and OPO Threshold Analysis}

To show that the combs inherit the topological properties of the linear system and are indeed confined to the boundary of the lattice, we perform direct imaging of the generated bulk and edge combs. While the system is designed to be well-confined in-plane, there is a certain amount of out-of-plane scattering caused by fabrication imperfections and disorder. The light scattered due to surface roughness is collected from above with a 10x objective lens and imaged on an infrared (IR) camera. In addition, we use a 1600 nm long-pass filter to remove the pump and only collect part of the generated comb light. 

Figure~\ref{Fig:osa}B shows the measured spatial intensity profiles of two types of generated frequency combs. The first row shows an edge comb. We observe that the generated comb light is
confined to the edge of the lattice and light travels from the input to the output port in the CW
direction (see the SI section S6 for simulated dynamics in different coupling regimes). Moreover, the
propagation is robust and no noticeable scattering into the bulk is observed from the two sharp
corners. These characteristics show that the comb teeth are indeed generated within the topological
edge band and that the topology is preserved even in the presence of strong nonlinearity. We note that the images shown here are taken from a copy of the main device that is presented throughout the main text. Comparison of the effect of different optical filters is shown in SI section S20. 

In sharp contrast to the edge comb, when we excite the lattice in the bulk bands, the spatial intensity distribution of generated comb light exhibits no confinement and occupies the bulk of the lattice. The result for a representative bulk comb is shown in the second row of Figure~\ref{Fig:osa}B. Additionally, Figure~\ref{Fig:osa}B shows simulated spatial distributions of edge and bulk modes in the linear regime for comparison. The details of nonlinear imaging and simulation for many bulk and edge combs are presented in the SI sections S6 and S7.

To study the OPO threshold for comb generation in our device, we perform pump power dependence measurement of the edge excitation. Figure~\ref{Fig:osa}C shows the edge comb pumped at 1547.8 nm wavelength. We observe an OPO threshold of around 80 mW average power. By increasing the pump power up to 210 mW, the triangular-like profile of the edge comb is maintained and widened. See the SI section S13 for comb power analysis as a function of pump wavelength including the bulk combs.

\subsubsection*{Nestedness of the comb and optical linewidth analysis}

To show the nested structure of the edge combs within each comb tooth, we measure the comb output at the drop port using a narrower band (1520nm to 1630 nm)  but ultra-high resolution (0.04 pm) heterodyne-based optical spectrum analyzer. The results for a few of the selected comb teeth are shown in Figure~\ref{Fig:osa}D. Within each of these comb teeth, we observe the oscillation of another set of well-resolved modes that correspond to the individual edge modes. The spacing between the oscillating edge modes is about 20 pm, which corresponds to the free spectral range of the super-ring $\nu_{\rm{S}}$ formed by the edge states and agrees with linear measurements. See the detailed demonstration of 20pm spacing in~\cite{flower2024observation} as well as SI section S21. Importantly, we observe the nestedness of the comb teeth throughout the edge band. Moreover, we observe that the bulk combs lack this nested structure. See SI sections S9 and S14 for the details of optical linewidth analysis and error estimation.

Next, we perform an optical linewidth analysis of the individual nested comb teeth, which provides insights on the noise level of the comb states~\cite{lei2022optical,zhang2019broadband}. We note that our pump has an optical linewidth (2.77 pm with an estimated error of $\pm 1.3$ pm, indicated as the red shaded area in Figure~\ref{Fig:osa}E, also see the SI sections S9 and S10 for details) that is above the resolution of our high-resolution (0.04 pm) OSA, which enables accurate monitoring of the variation of the optical linewidth of the edge and bulk combs compared to the pump throughout this work. Figure~\ref{Fig:osa}E shows the linewidth of the edge modes for the longitudinal mode $\mu = 1$ when the pump wavelength is set near the center of the edge band at 1547.82 nm. We observe around an order of magnitude linewidth reduction for the comb tooth linewidth (blue circles) compared to the cold-cavity counterparts (red circles) which were obtained from the linear drop port transmission measurements. The red dashed line indicates the pump's optical linewidth. We note that the comb tooth linewidth is close and at times lower than the pump's linewidth. 

Moreover, we perform the optical linewidth analysis for the full pump wavelength range for an individual edge mode. The results for the nested tooth $\sigma = 2$ (wavelength 1553.845 nm) are shown in Figure~\ref{Fig:esa}F. We observe that throughout the edge band, the nested tooth's linewidth is close to the pump's linewidth, and it diverges as the pump is detuned to the bulk on either side. Importantly, this observation indicates that the narrowness of the optical linewidths of the edge combs is predominantly set by the pump's linewidth. We note that the competition between different noise sources can sometimes lead to comb tooth linewidths becoming even narrower than the pump to a minimum value known as the ``fixed point"~\cite{lei2022optical}, however, the error values in our data are not sufficiently low to meticulously investigate this effect. Detailed analysis of nesting across the entire wavelength range and over different teeth is presented in the SI section S8.

\subsection*{Signatures of mode-locking in slow and fast timescales}

We study the mode-locking signatures of our nested combs with three independent methods. For simultaneous mode-locking in the fast and slow timescales, an optical linewidth analysis of the nested comb lines is used. Separately, since the slow timescale of our system is in the few GHz regime, this makes it possible to investigate the signatures of slow timescale mode-locking directly with electronic noise analysis of the comb states as well as their time-domain dynamics.

\subsubsection*{Simultaneous mode-locking in fast and slow timescales}

Unlike the slow super-ring time scale of 2.5 GHz, direct detection of signatures of the fast time scale of the single-ring round trip time (750 GHz) is beyond the capability of typical oscilloscopes and ESAs. Nevertheless, the elastic tape model~\cite{lei2022optical,telle2002kerr} provides an accurate theoretical description of optical signatures of mode-locked microcombs that can be measured via optical linewidth analysis. Specifically, this model which is an ab-initio derivation assuming a fixed repetition rate for a train of pulses, shows the comb tooth linewidth increases quadratically as a function of mode number separation with respect to a fixed point ~\cite{lei2022optical}. 

To investigate the existence of such noise-analysis signatures of the mode-locking in the both fast and slow timescales of our system, we analyzed the optical linewidth of nested comb lines across several longitudinal modes of the single rings $\mu$. The results for a typical edge comb (here shown for nested tooth $\sigma = 3$) are shown in Figure~\ref{Fig:esa}A. We clearly observe a quadratic form of the optical linewidth, which goes as low as the pump's linewidth near the pumped mode and increases quadratically for modes away from the pump. Remarkably, in the same measurement, we observe this quadratic variation in the comb tooth linewidth with respect to the minimum linewidth near the pumped mode simultaneously over the slow timescale. We show this in the inset of Figure~\ref{Fig:esa}A, where we plot the optical linewidth of the corresponding nested comb for typical longitudinal modes (here for $\mu = -1$ and $\mu = 1$). This remarkable simultaneous quadraticity of the variation in the comb tooth linewidth over the two timescales of the edge combs in our system is clearly observed over a wide range of pump wavelengths within the edge band and for many longitudinal modes $\mu$, the data for which are presented in the SI section S14.

Moreover, we observe the quadratic form of the optical linewidth only for the edge combs, and the bulk counterparts lack such features, as shown in Figure~\ref{Fig:esa}B for a typical bulk comb (pump wavelength 1547.91 nm). We note that since the bulk combs typically lack the nested structure, the linewidth analysis presented in Figure~\ref{Fig:esa}E is only possible for the fast timescale. Further, as another direct comparison with edge combs, we performed the same optical linewidth analysis for modulation-instability combs generated in a comparable single race-track device and observed the absence of nestedness and much higher optical linewidth compared to the pump, in sharp contrast to our topological combs. Complementary analyses of this study and bulk combs are available in the SI section S14. 

\subsubsection*{Mode-locking in the slow timescale}

We perform electrical noise analysis of the comb states with a fast (43.5 GHz) ESA. We note that in comparison to the oscilloscope, our ESA provides around two orders of magnitude larger dynamic range ($\approx$ 100 dB), allowing for a finer study of low-noise comb states.

\begin{figure*}[t]
    \centering
    \includegraphics[width=0.8\textwidth]{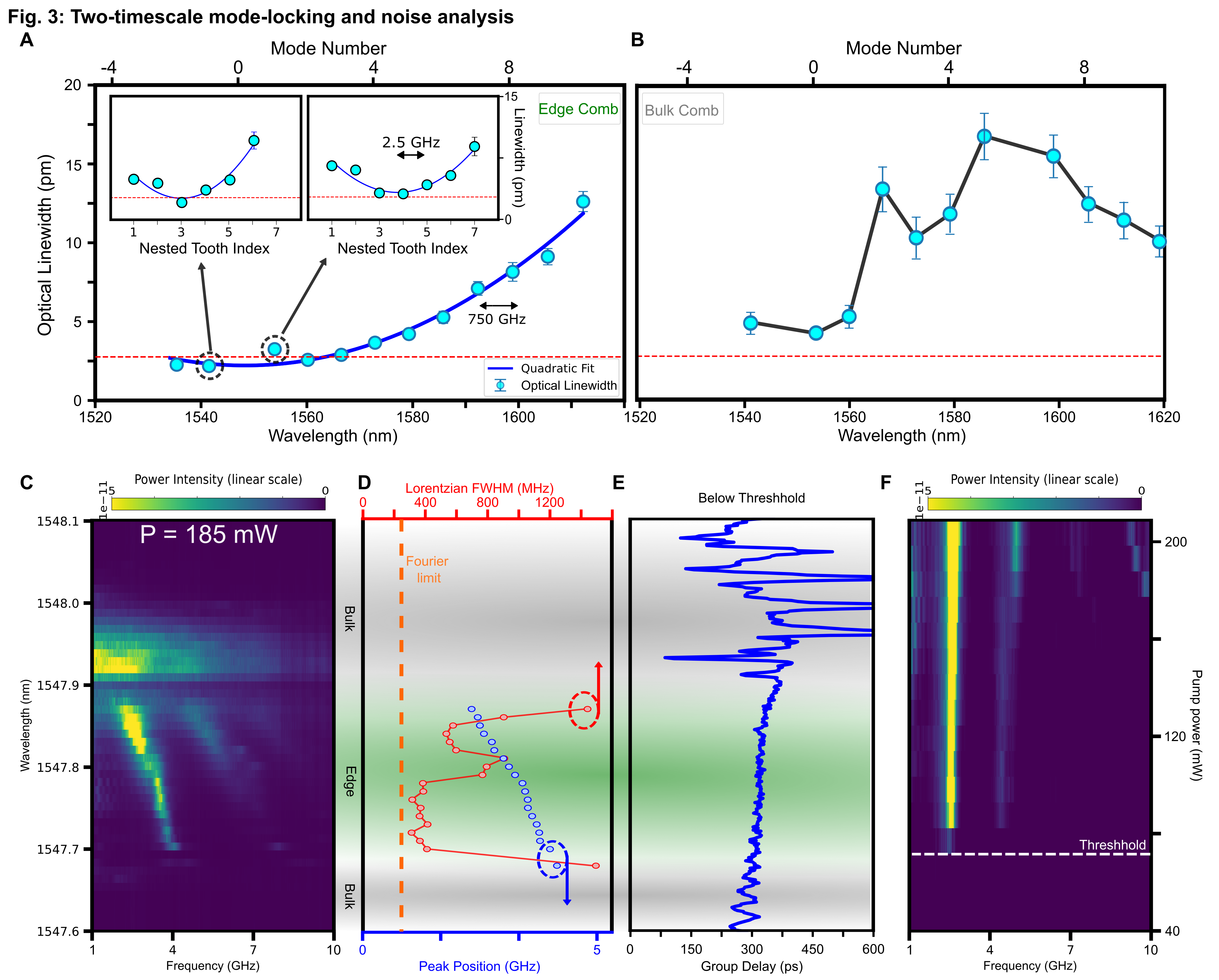}
    \caption{\textbf{Two-timescale mode-locking and noise analysis. }\textbf{A,} Quadratic variation of the comb teeth optical linewidth for an edge comb with a pump wavelength of 1547.86 nm (nested tooth $\sigma = 3$), across several longitudinal modes of the fast (750 GHz) and slow (2.5 GHz) timescales, indicated by $\mu$ and $\sigma$, respectively. The data for the slow timescale shown in the inset are for (left) $\mu = -1$ and (right) $\mu = 1$. The blue curve and the red dashed line are the quadratic fit and the pump's linewidth, respectively. \textbf{B,} The optical linewidth variation over the fast timescale for a typical bulk comb (pump wavelength 1547.91 nm), which lacks the quadratic form of the edge counterparts. The shaded red area in \textbf{A,B,} indicates the estimated error for the pump linewidth (see Figure~\ref{Fig:optical_linewidth} for details). \textbf{C,} Electrical spectrum analysis (ESA) of the topological combs as a function of pump wavelength with $\approx$ 185 mW of average pump power. The bulk and edge regions of the spectrum are highlighted with gray and green, respectively. \textbf{D,} Peak position (blue) and Lorentzian linewidth analysis (red) of the repetition rates as a function of pump wavelength. The orange dashed line is the Fourier transform limit set by the 5 ns pulse duration of the pump (see SI section S11 for details). Note that here the uncertainties are smaller than the symbols. \textbf{E,} Group delay measurement of the lattice in the linear regime using a weak tunable laser. \textbf{F,} Repetition beats of an edge comb as a function of pump power. The white dashed line indicates the OPO threshold.}
    \label{Fig:esa}
\end{figure*}

We perform the noise analysis by sending the combs (after filtering the pump) to the PD and analyzing the RF output with the ESA. Figure~\ref{Fig:esa}C shows the results as a function of pump wavelength (generated with a constant 185 mW of pump power). Importantly, well-defined repetition beats are observed only in the edge band in contrast to the bulk bands. Moreover, the observed repetition rates of the combs are in good agreement with the time-domain observations shown later in Figure~\ref{Fig:scope}G. 

Importantly, the high dynamic range of the ESA enables an accurate linewidth analysis of the combs' beats and their noise level. We obtain this by performing Lorentzian fitting of the frequency beats as a function of pump wavelength. Figure~\ref{Fig:esa}D shows the repetition rates (peak's central positions shown in blue circles) as well as their corresponding linewidth (red circles). The vertical orange dashed line indicates the 223 MHz Fourier transform (FT) limit which is set on the frequency resolution of the measurement due to the finite 5 ns pulse duration of the pump. A detailed analysis of this limit is presented in the SI section S11.

We note two important observations in Figure~\ref{Fig:esa}B. First, the linewidth of the ESA beat notes within the edge band approaches the transform limit, which highly suggests the existence of mode-locked comb states. Moreover, this near-transform-limited linewidth persists over a relatively wide portion of the edge band, suggesting a broad range of possible mode-locking states. Importantly, linewidths start to diverge as the pump is detuned to the bulk bands on either side of the edge bands. This is in close agreement with the time domain results shown in Figure~\ref{Fig:scope}G. Second, the repetition rates within the edge band vary from 2.5 GHz to 4 GHz. This observation, in part, is understood using the group delay measurement of the device in the linear regime. As shown in Figure~\ref{Fig:esa}E, the edge part of the spectrum shows an increase of the round trip time between 300-400 ps, when the pump is detuned from the shorter to longer wavelength part of the edge band. This $\approx$ 25~$\%$ change in the round trip and its increasing trend is the dominant contributor to the observed $\approx$ 
 35~$\%$ reduction of the repetition rate seen in Figure~\ref{Fig:esa}B. We note that the bulk has no well-defined round trip time which indicates the absence of well-defined repetition rates of the bulk combs.

For an edge comb, we also perform the power dependence measurement of the ESA study. As shown in Figure~\ref{Fig:esa}D, above the OPO threshold the edge comb repetition rates can be seen, and their higher harmonics also appear at higher pump powers.

\subsubsection*{Temporal signatures of mode-locking}

Since the slow timescale of our device is designed to be a few GHz, this enables direct study of the temporal dynamics and pattern formation of the bulk and edge combs with a fast oscilloscope. We obtained this by sending the entire optical signal of the combs (excluding the pump) to a fast (50 GHz) PD and processing the RF output with a fast (20 GHz) oscilloscope. The results are shown in Figure~\ref{Fig:scope}.

We first temporally characterize our pump by tuning the pump into the center of the edge band (1547.8 nm), coupling 40 mW of peak power into the input port of the device, and sending the drop port's output (without filtering the pump) to the PD. Figure~\ref{Fig:scope}A,D shows the PD's output detected by the oscilloscope and their FT, respectively. The plots include 2000 repeats of the experiments, which shows the pump's jitter over time. A representative example (blue) and the average (red) are also shown. Since 40 mW lies well below the OPO threshold ($\approx$ 80 mW) of our device, at this pump power simply the 5 ns pulse of the pump is observed at the drop port as expected. Importantly, we also observe a variation of the pulses between different experiments which indicates a certain amount of pump jitter. The role of this jitter in our comb generation is discussed later.

Next, we increase the pump power above the OPO threshold at the same edge band wavelength with 185 mW of peak power and generate the combs. Figure~\ref{Fig:scope}B,E shows the PD's output detected by the oscilloscope (here the pump is filtered out) and their FT, respectively. Here we clearly observed the formation of periodic temporal patterns within the 5 ns pulse duration of the pump. Importantly, the observed periodicity of 300 ps matches the designed round trip time of our topological device at this pump wavelength. The difference between the individual pulses (blue) and the average (red) originates partly from the pump jitter, as shown in Figure~\ref{Fig:scope}A. The FT analysis of the 2000 repeated experiments shown in Figure~\ref{Fig:scope}E clearly demonstrates a persistent repetition rate of the combs which for this pump wavelength is at $\approx$ 3.3 GHz.

To compare the temporal dynamics of the edge combs with their bulk counterparts, we perform the same measurement by changing the pump wavelength into the bulk at 1547.9 nm. As shown in Figure~\ref{Fig:scope}C,F, no corresponding dominant Fourier frequency in the electrical spectrum is observed, which originates from the lack of well-defined round trip time of the bulk states in contrast to the edge counterparts.

To comprehensively study the temporal dynamics in the bulk versus the edge, we perform the same measurement for a wide range of pump wavelengths, starting from the bulk at the shorter-wavelength part of the spectrum (1547.6 nm) up to the longer-wavelength part (1548.1 nm), covering the entire edge band. The FTs of the results are shown in Figure~\ref{Fig:scope}G. The spaced vertical panels are measured at 30 pm intervals. Importantly, we observed that well-defined repetition rates emerged as the pump detuned from the bulk on either side towards the center of the edge band.

Next, we investigate the pump power dependence of the temporal patterns. Figure~\ref{Fig:scope}H shows the direct time-domain results for an edge (1547.8 nm) comb. The formation of temporal patterns above the OPO threshold can be seen clearly, which fills the 5 ns duration of the pump pulse at higher pump powers. The same measurement for the bulk counterpart shows no temporal periodicity emerging. A comprehensive study of temporal pattern formation in edge versus bulk excitation and their pump power dependence is presented in the SI sections S15-S18.   

\section*{Discussion}

Here for the first time, we demonstrated signatures of topological spatiotemporal mode-locking in a multi-timescale 2D array of hundreds of coupled ring resonators. We further provide results for a second copy of the device with different parts of the comb, establishing the repeatability of the experiment and strengthening our observation, as shown in SI section S20. We note that due to limitations of our experiment as shown in SI section S22, further studies are needed to confirm the existence of topological solitons~\cite{mittal2021topological}. Nevertheless, our results indicate topological mode-locking and non-harmonic multi-timescale synchronization between photonic modes of hundreds of optical resonators is possible in a compact chip architecture. Moreover, we demonstrated clear spatial, optical, and electrical noise signatures for the identification and classification of different types of edge and bulk states, which not only adds numerous understanding to the preceding work, but also constitutes a way forward for understanding complex nonlinear states of light in topological photonics. Going forward, the uncoupled nature of the two timescales in our mode-locking scheme opens the possibility of independently tuning the timescales into a variety of configurations~\cite{dai2024programmable,ma2024anisotropic}. Moreover, the nesting in our system can be cascaded further by adding another lattice layer with an independent coupling rate, opening the door for non-harmonic many-timescale mode-locking from MHz to THz, as well as exploring nonlinear multi-mode optical synchronization phenomena~\cite{wright2022nonlinear}.

On a more fundamental level, our results provide an innovative platform to study the interplay of topology and optical nonlinearities~\cite{smirnova2020nonlinear,jurgensen2021quantized,mostaan2022quantized,mittal2021topological}, as well as intriguing nonlinear topological physics unique to bosons~\cite{szameit2024discrete,smirnova2020nonlinear}, such as nonlinearity-induced restructuring of the bulk-edge correspondence~\cite{Maczewsky2020,Kirsch2021} and nonlinear control of topology~\cite{xia2021nonlinear,maczewsky2020nonlinearity,dai2024non,sone2024nonlinearity}. Looking forward, beyond our AQH lattice, intriguing avenues open up to explore nonlinear optical synchronization phenomena in the presence of other synthetic gauge fields, such as Floquet~\cite{hashemi2024floquet} and integer Hall models, other topological models~\cite{ozawa2019topological,jalali2023topological} and frequency regimes~\cite{sharp2024near}, and Kerr-induced symmetry-breaking effects that are unique to coupled-resonator systems~\cite{ghosh2024controlled,sanyal2024nonlinear,tusnin2023nonlinear}. Moreover, the nested character of our combs provides a higher-dimensional testbed for exploring comb spectroscopy over different precision scales, as well as fundamentals of noise physics in coherent solitonic solutions~\cite{lei2022optical}. Further, our platform may be adapted to study topology in other mode-locked photonic sources, including synchronously pumped OPOs~\cite{marandi2014network,honjo2021100,roy2022temporal}, topological temporally mode-locked lasers~\cite{leefmans2024topological}, and dissipative Kerr cavities~\cite{englebert2021temporal}. The gradual tunability of our edge-to-bulk combs also can facilitate the investigation of theoretically proposed chimera-like states -coexisting chaotic and coherent states of the electromagnetic field inside the cavity~\cite{tusnin2020nonlinear}.


\section*{ Materials and Methods}

\subsection*{Device fabrication}

The topological photonic devices were fabricated in a commercial foundry in the same way as the ones presented in Ref.~\cite{flower2024observation}. A high-resolution optical image presented in Figure~\ref{Fig:band_Linear} shows the topological photonic lattice used in this work. The microring resonators are made of Si$_3$N$_4$ embedded in SiO$_2$. Their dimensions are 1200 nm wide and 800 nm thick. The gaps between the microrings, as well as those between the bus waveguides and the microrings, are 300 nm.

\subsection*{Experimental setup}

For the linear measurements as well as the nonlinear imaging, we closely followed the method presented in Ref.~\cite{flower2024observation}.

For temporal measurement and electrical noise analysis, we first sent the combs to a fast (50 GHz) PD and analyzed the RF output with a fast (43.5 GHz) ESA as well as a fast (20 GHz) oscilloscope. For these measurements, the pump was removed from the spectrum prior to the PD using a notch filter (see SI section S12 for details of the filter).

\begin{figure*}[t]
    \centering
    \includegraphics[width=1.0\textwidth]{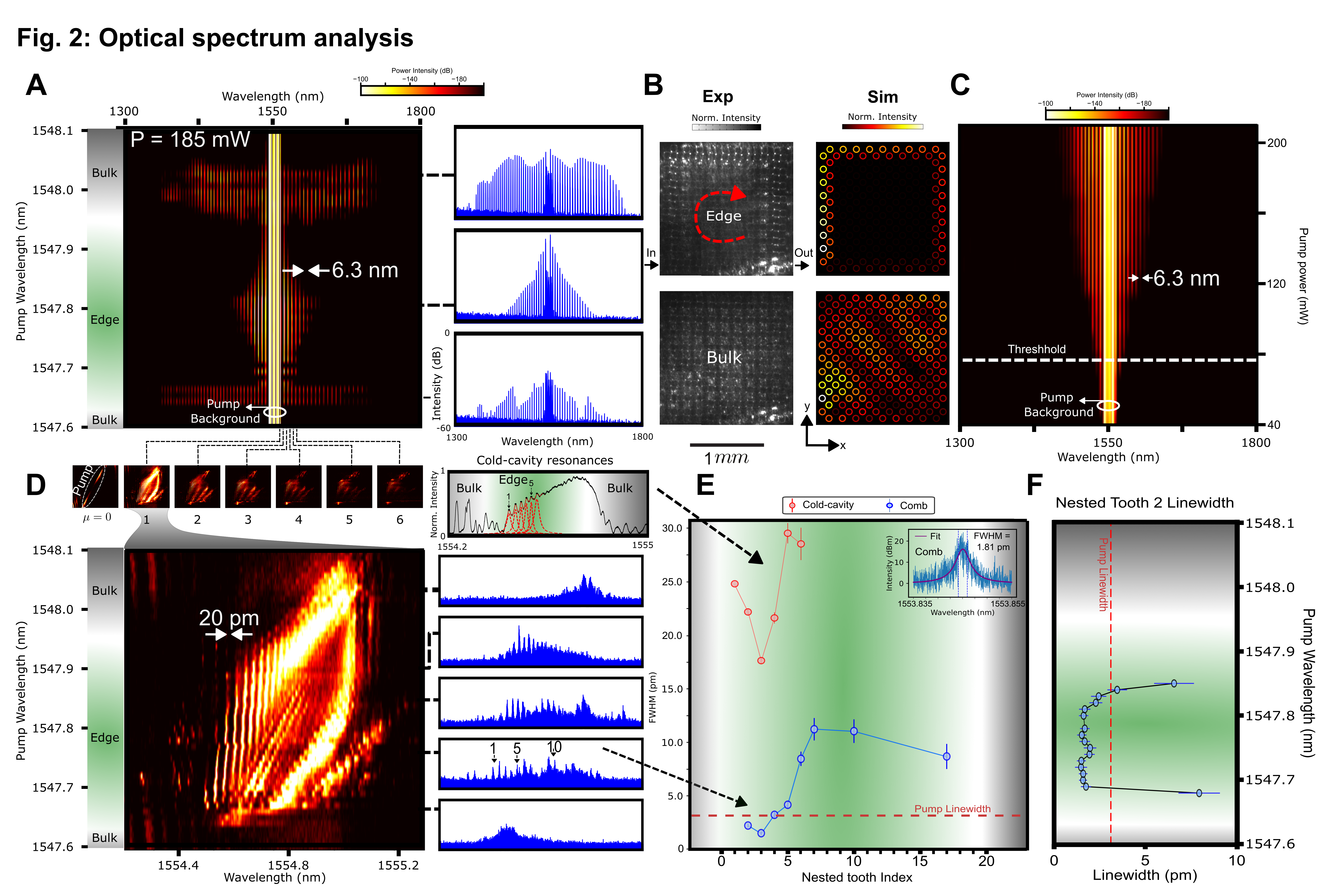}
    \caption{\textbf{Optical spectrum analysis. }\textbf{A,} Broadband optical spectrum of topological combs vs. pump wavelength at $\sim$185 mW average pump power. Bulk and edge regions are shaded gray and green, respectively. Three representative snapshots are shown on the right. 
\textbf{B,} (Left) Measured spatial intensity profiles of bulk and edge combs; (right) corresponding linear simulations with only clockwise excitation. The device is the same design as in \textbf{A} but exhibits higher local scattering at the damaged output port. 
\textbf{C,} OSA spectrum of an edge comb vs. average pump power. The white dashed line marks the OPO threshold. Color scale is in dBm (0 dB = 1 mW). 
\textbf{D,} High-resolution spectra of individual comb teeth ($\mu$) vs. pump wavelength. Edge and bulk bands are shaded green and gray, respectively. Five snapshots and cold-cavity resonances are included. 
\textbf{E,} Linewidths of nested longitudinal edge modes ($\sigma$), with cold-cavity linewidths shown for comparison. Inset: example fit of a nested tooth. 
\textbf{F,} Optical linewidth of the $\sigma = 2$ tooth vs. pump wavelength. Red shading in \textbf{E,F} indicates estimated pump linewidth uncertainty, arising from jitter and OSA acquisition time (see Figure S6).
}
    \label{Fig:osa}
\end{figure*}

\begin{figure*}[t]
    \centering
    \includegraphics[width=0.99\textwidth]{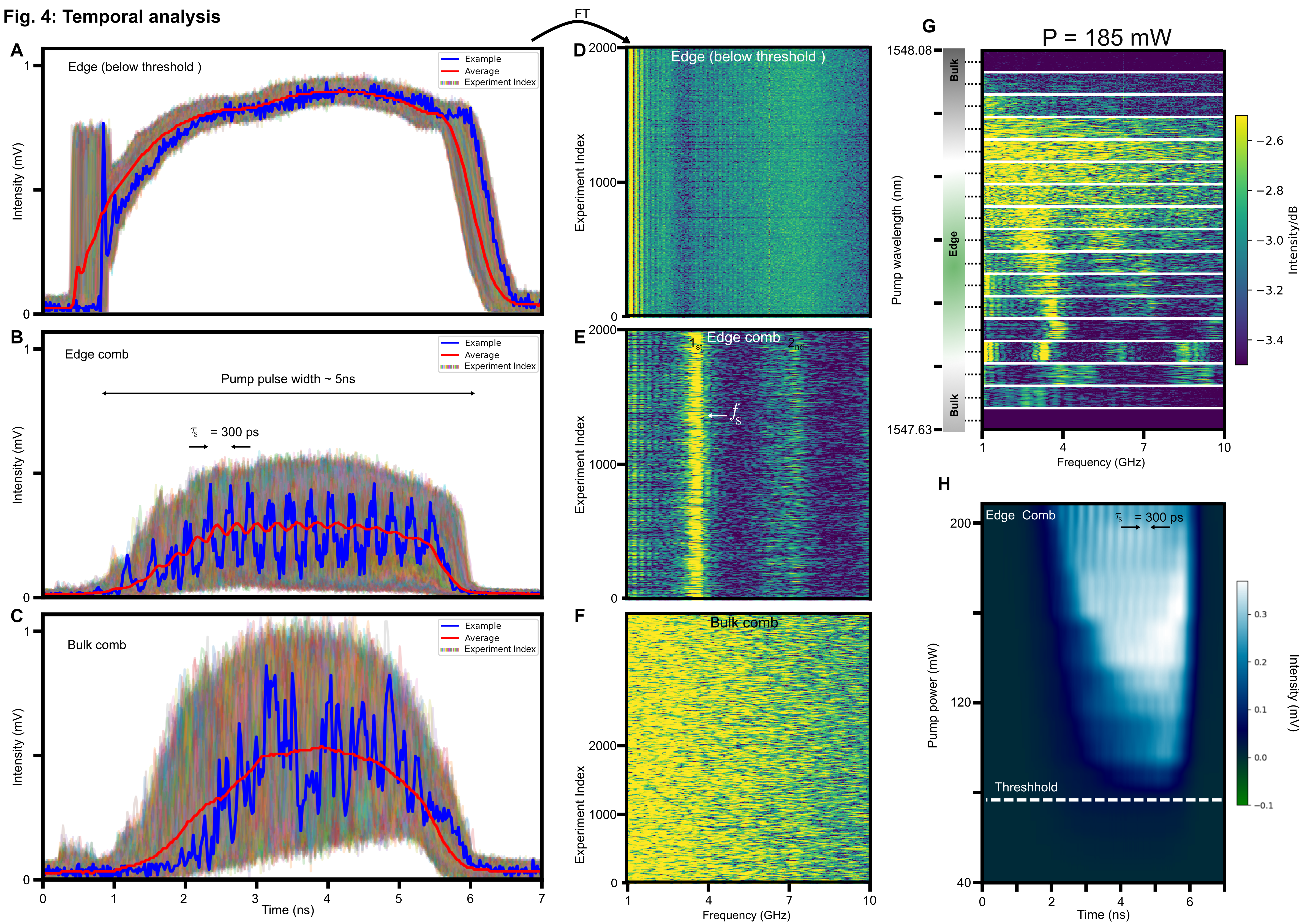}
    \caption{\textbf{Temporal measurement of the device}. When the output at the drop port is sent to a fast PD, with \textbf{A,} approximately 40 mW of on-chip peak pump power, which is well below the OPO threshold. Pumping the \textbf{B,} edge and \textbf{C,} bulk above the OPO threshold with 185 mW pump power. The plots include 2000 repeats of the experiment, with one example shown in blue and the average shown in red. \textbf{D-F,} corresponding Fourier transformations (FTs) of the data in \textbf{A-C,}, respectively. \textbf{G,} FT of the oscilloscope measurements of the topological combs as a function of discreet pump wavelength taken at every 30 pm intervals with approximately 185 mW of on-chip peak pump power. The bulk and edge regions of the spectrum are marked. Each vertical strip includes 2000 repeats of the experiment. \textbf{H,} Oscilloscope measurement of an edge comb as a function of pump power. The white dashed line indicates the OPO threshold.
    }
    \label{Fig:scope}
\end{figure*}

%
%
%
%
%
%


\section*{Acknowledgements}

The authors wish to acknowledge fruitful discussions with Apurva Padhye, Deric Session, Avik Dutt, and Daniel Leykam. 

\paragraph*{Funding}
This work was supported by ONR N00014-20-1-2325, and MURI FA9550-22-1-0339.

\paragraph*{Authors contribution}
L.X. and M.J.M. performed the experiments and simulations and analyzed the data. C.J.F. designed the devices and contributed to the early stages of the measurements and data analysis. C.J.F., L.X., M.J.M., A.R., and D.G.S.F. contributed to the construction of the experimental setup. M.J.M. wrote the manuscript. The project was supervised by Y.C., S.M., G.M., K.S., and M.H. All authors discussed the results and contributed to the manuscript.

\paragraph*{Competing interests}
S.M. and M.H. are inventors on a patent (US Patent 11599006) dated 7 March 2023 that covers the generation of nested frequency combs in a topological source. The authors declare no other competing interests.

\paragraph*{Data availability}
All of the data that support the findings of this study are reported in the main text and Supplementary Information. Source data are available at \href{https://zenodo.org/records/15492780}{https://zenodo.org/records/15492780}.

\subsection*{Supplementary materials}

Supplementary Text\\
Figs. S1 to S24\\
Movie S1\\


\newpage


\renewcommand{\thefigure}{S\arabic{figure}}
\renewcommand{\thetable}{S\arabic{table}}
\renewcommand{\theequation}{S\arabic{equation}}
\renewcommand{\thepage}{S\arabic{page}}
\setcounter{figure}{0}
\setcounter{table}{0}
\setcounter{equation}{0}
\setcounter{page}{1} 


\begin{center}
\section*{Supplementary Materials for\\ \scititle}

Lida~Xu$^{1\dagger}$, Mahmoud Jalali Mehrabad$^{1\ast\dagger}$, Christopher J. Flower$^{1\dagger}$\\
Gregory Moille$^{2}$, Alessandro Restelli$^{2}$, Daniel G. Suarez-Forero$^{1}$, Yanne Chembo$^{3}$\\ Sunil Mittal$^{4}$, Kartik Srinivasan$^{2}$, Mohammad Hafezi$^{1\ast}$\\
\small$^\ast$Corresponding author. Email: mjalalim@umd.edu, hafezi@umd.edu\\
\small$^\dagger$These authors contributed equally to this work.
\end{center}

\subsubsection*{This PDF file includes:}
Supplementary Text\\
Figures S1 to S24\\

\subsubsection*{Other Supplementary Materials for this manuscript:}
Movie S1 

\newpage


\subsection*{\label{si:design} S1: Design and Hamiltonian description}
 
The device design and Hamiltonian description of our system, including the linear and nonlinear dynamics as well as drive and dissipation, closely follow the device described in Ref.~\cite{flower2024observation}. 


\subsection*{ S2: Device parameter estimation}

A high-resolution optical image in \ref{Fig:band_Linear}c illustrates the topological photonic lattice utilized in this study. The device comprises a two-dimensional array of 261 interconnected photonic ring resonators with a pair of coupled input-output waveguides. These waveguides, embedded in silicon dioxide, have dimensions of 1200 nm in width and 800 nm in thickness to ensure operation within the anomalous dispersion regime. Simulated mode profiles and dispersion characteristics are presented in Figure~\ref{Fig:lin_sims} of the Supplementary Material of Ref.~\cite{LukeSiN}. Each ring features a racetrack structure, incorporating $12$ $\mu$m straight coupling sections and $90^{\circ}$ Euler bend regions, with an effective radius of $20$ $\mu$m, resulting in a free spectral range (FSR) of approximately $0.75$ THz.

The coupling gaps between the resonators, as well as those between the input-output waveguides and the resonators, are set to 300 nm, yielding an approximate coupling strength, $J$, of $2\pi\times25$ GHz. The extrinsic and intrinsic couplings, denoted as $\kappa_{\rm ex}$ and $\kappa_{\rm in}$, are estimated at $2\pi\times30$ GHz and $2\pi\times2$ GHz, respectively (see Figure~\ref{Fig:lin_sims}). Further details on these calculations can be found in the Supplementary Material of Ref.~\cite{flower2024observation}.

The fast timescale (1 THz) comes from the length \textit{L}, the material, and geometry of the single rings that construct the lattice. Consider an ideal circular ring resonator with group refractive index $n_g$, denoting speed of light as $c$, the Free Spectral Range (FSR) for a single ring is $c/n_gL$. The slow timescale corresponds to the spacing between two adjacent super-ring modes in the edge band, as shown in Figure~\ref{Fig:lin_sims}. The coupling strength \textit{J} between two adjacent rings is controlled by their gap. The size of the lattice is described by the number of rings \textit{N} on one side. The whole edge band spans \textit{2J}, and there are N edge modes with linear dispersion inside the edge band, giving us a slow timescale $2J/N$. Since lattice parameters are independent of single ring parameters, the fast and slow timescales are independently designable.


\subsection*{S3: Band structure}
Figure~\ref{Fig:band_Linear}a-b presents the unit cell and the calculated band structure of a semi-infinite AQH lattice, periodic along one axis and finite along the other. A $2J$-wide edge band region, marked in green, lies between two bulk bands marked in gray. In contrast to the bulk modes, which do not possess well-defined momentum, two unidirectional edge bands with opposite pseudospins exist. These edge states propagate in opposite directions and exhibit robustness against local disorder~\cite{leykam2018reconfigurable,mittal2019photonic}.

 
\subsection*{S4: Linear characterization}
\begin{figure*}[t]
     \centering     \includegraphics[width=0.96\textwidth]{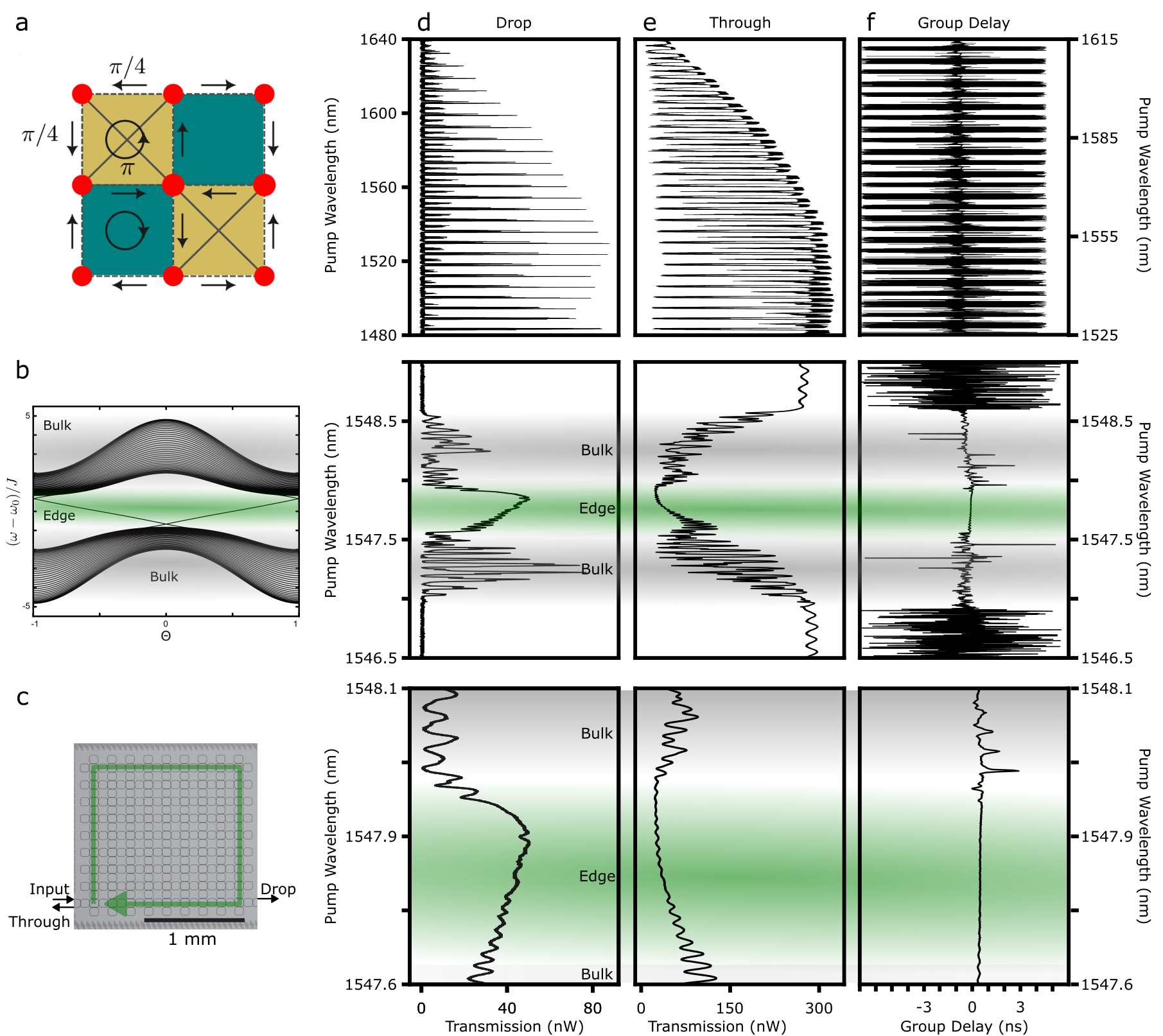}
     \caption{\textbf{Band structure and linear characterization} (a) A schematic of the unit cell of the lattice. (b) The band structure of a semi-infinite lattice (finite along the y-axis, periodic boundary conditions along the x-axis). Here $\Theta$ is the phase (normalized by $\pi$) between neighboring site-rings along the axis with periodic boundary conditions. The edge and bulk bands are highlighted in green and gray, respectively. (c) High-resolution optical image of the AQH device. Measured (d) drop and (e) through port transmission spectrum of the topological lattice showing bulk and edge bands, for many (top row) and one zoomed individual (middle row) modes. The bottom row shows the pump wavelength range used for all the nonlinear measurements. (f) The corresponding group delay spectrum of (d-e) showing a flat edge band, measured at the drop port of the device. } 
     \label{Fig:band_Linear}
 \end{figure*}

This section closely follows the method described in Ref.~\cite{flower2024observation}. A high-resolution optical image of the AQH lattice is shown in Figure~\ref{Fig:band_Linear}c. The linear characterization (using weak continuous-wave tunable lasers) of the device is shown in Figure~\ref{Fig:band_Linear}d-e. This includes the device's output measured from the drop and through ports over many longitudinal (top row) as well as individual (middle row) modes, in addition to the pump wavelength range used during all the nonlinear measurements (bottom row). Moreover, Figure~\ref{Fig:band_Linear}f shows the corresponding linear-regime group delay results of Figure~\ref{Fig:band_Linear}d-e measured from the drop port of the device analyzed with a vector analyzer.   

\subsection*{ S5: Experimental setup}

In linear regime measurements, a continuous-wave tunable laser is coupled into the input port using edge couplers, and the wavelength is continuously swept. Simultaneously, the output power at the drop port across the lattice is measured using a power meter. To determine the group delay of the transmission at the drop port, an optical vector network analyzer is utilized. Polarization control in both setups is managed using a standard 3-paddle polarization controller located at the laser output.

\begin{figure*}[t]
     \centering
     \includegraphics[width=0.99\textwidth]{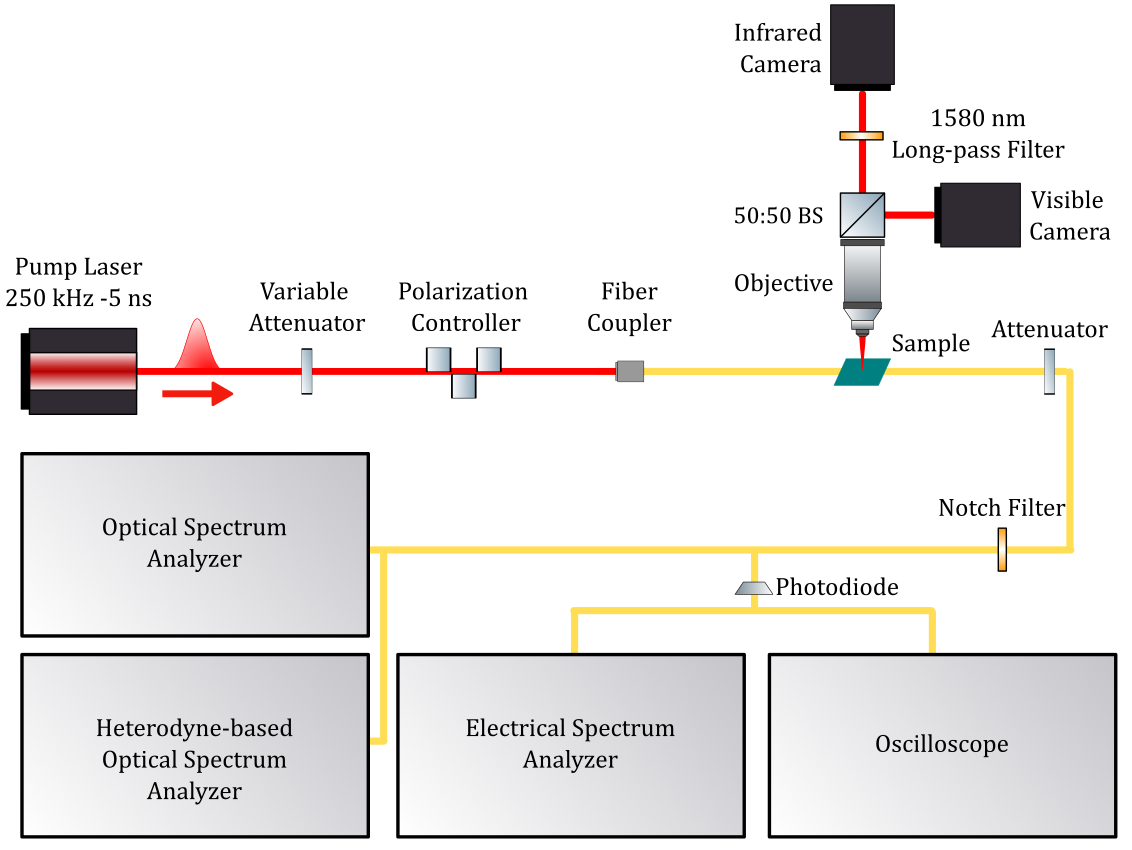}
     \caption{\textbf{Detailed schematic of the measurement setup.} A tunable telecom pulsed laser is sent through a variable attenuator and polarization controller before being fiber-coupled and sent into the SiN device. The output of the device is then fiber-coupled and optionally attenuated by a second variable attenuator and notch filter for pump removal before being sent to the grating-based and heterodyne-based OSAs, ESA, and oscilloscope. The chip is also imaged from above with a 10x objective, followed by a 50:50 beamsplitter. One optical path is sent to a visible camera, while the other is filtered by a 1580 nm long-pass filter and sent to an IR-sensitive camera. The collected comb from the output coupler is then sent for optical, electrical, and temporal analysis. BS: beam splitter
     } 
     \label{Fig:exp_schematic}
 \end{figure*}

Figure~\ref{Fig:exp_schematic} also provides a schematic of the experimental setup used for nonlinear measurements. In this setup, a pulsed tunable laser is directed into a free-space optical arrangement, which includes a variable attenuator and a polarization controller made up of a quarter-wave plate, a half-wave plate, and another quarter-wave plate. The output is subsequently coupled into a short tapered fiber and edge-coupled into the SiN chip via the input port. Coupling losses for each coupler are estimated at 2 dB to 3 dB. The combined pump and comb output is collected from the drop port using another tapered fiber and can be optionally attenuated or filtered before being analyzed. The output is then directed to broadband low-resolution (40 pm) and high-resolution (0.4 pm) optical spectrum analyzers (OSAs) based on gratings and heterodyne detection, respectively, as well as to an electrical spectrum analyzer (ESA) and an oscilloscope.

For imaging, out-of-plane scattering from the SiN chip is captured using a 10x objective lens with a numerical aperture of 0.28. The image is directed through a 50:50 beamsplitter, allowing it to be captured by both a visible wavelength camera and an infrared (IR) sensitive camera. A 1580 nm long-pass filter is positioned in front of the IR camera to block the pump laser.

For temporal measurements and electrical noise analysis, the comb is directed to a fast (50 GHz) PD. The resulting RF output is then analyzed using a fast (43.5 GHz) ESA and a high-speed (20 GHz) oscilloscope.


%
\subsection*{  S6: Linear simulations}

\begin{figure*}[t]
     \centering     \includegraphics[width=0.99\textwidth]{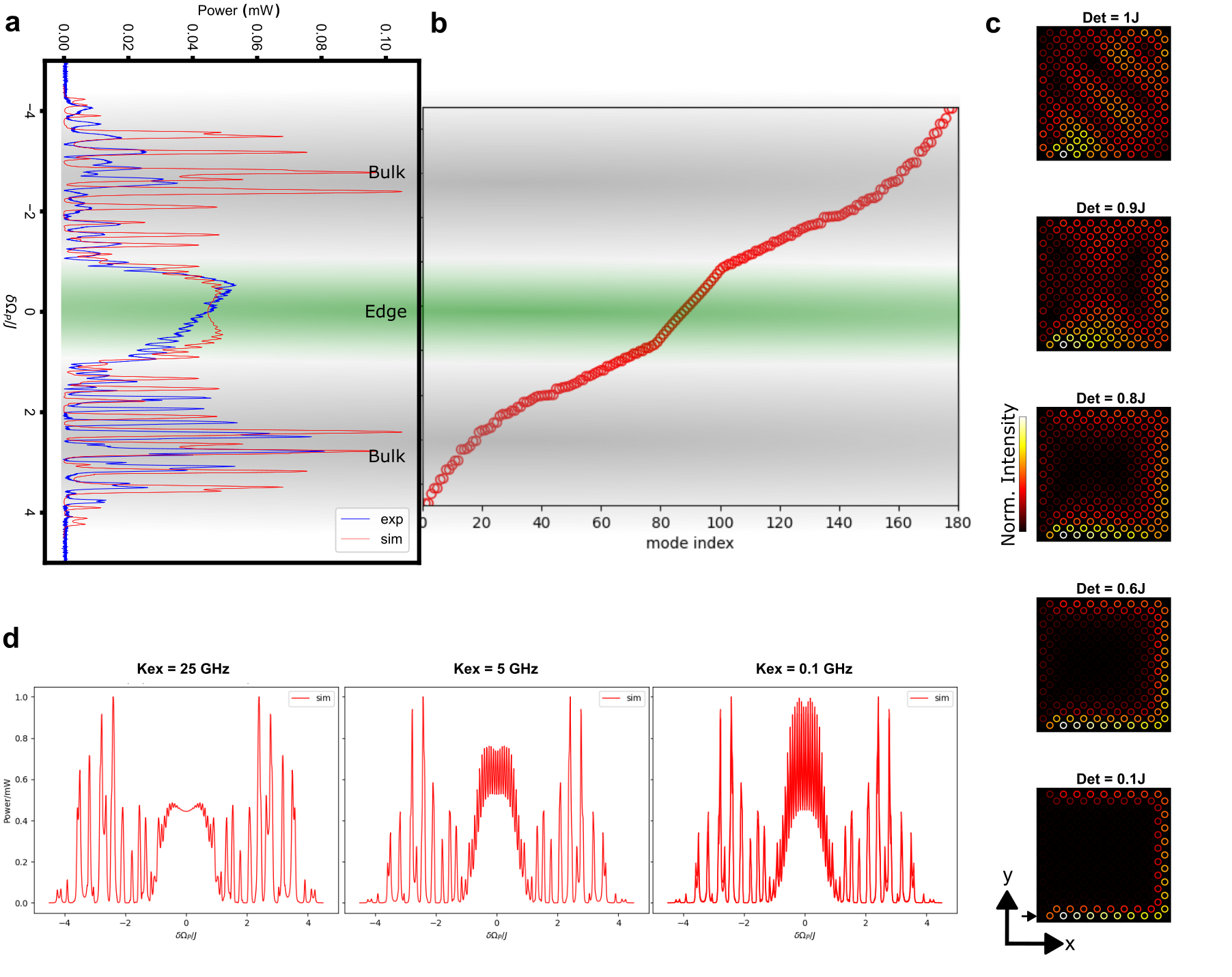}
     \caption{\textbf{Linear Simulations
     } (a) Simulated (red) and measured (blue) linear-regime (weak pump power) transmission spectrum at the drop port of the device for a single longitudinal mode. Calculated eigenvalues (b) and selected spatial intensity profiles (c) of the device in the linear regime. (d) Simulated dependency of the linear transmission spectrum on the $\kappa_{ex}$.}

     \label{Fig:lin_sims}
 \end{figure*}

Based on the linear Hamiltonian description discussed in Ref.~\cite{flower2024observation}, the single-mode linear simulations of the drop port transmission spectrum and corresponding experimental data of our device are shown in Figure~\ref{Fig:lin_sims}a. Figure~\ref{Fig:lin_sims}b-c shows the corresponding eigenvalues as well as selected spatial intensity profiles of a few eigenmodes. Note that here the output ring is located upper left. It can be clearly seen that the degree of confinement of the edge modes decreases by moving the pump frequency away from the center of the edge band, eventually entering the bulk band with distinct bulk profiles.

Moreover, as discussed earlier, we note that $\kappa_{ex}$ in our device is designed to be 25 GHz, which leads to the broadening and merging of edge resonance within the edge band. Figure~\ref{Fig:lin_sims}d shows how by reducing  $\kappa_{ex}$ to smaller values (here calculated for a fixed $J$ and $\kappa_{in}$), the edge modes can be spectrally resolved.

\subsection*{  S7: Nonlinear imaging}

\begin{figure*}[t]
     \centering     \includegraphics[width=0.9\textwidth]{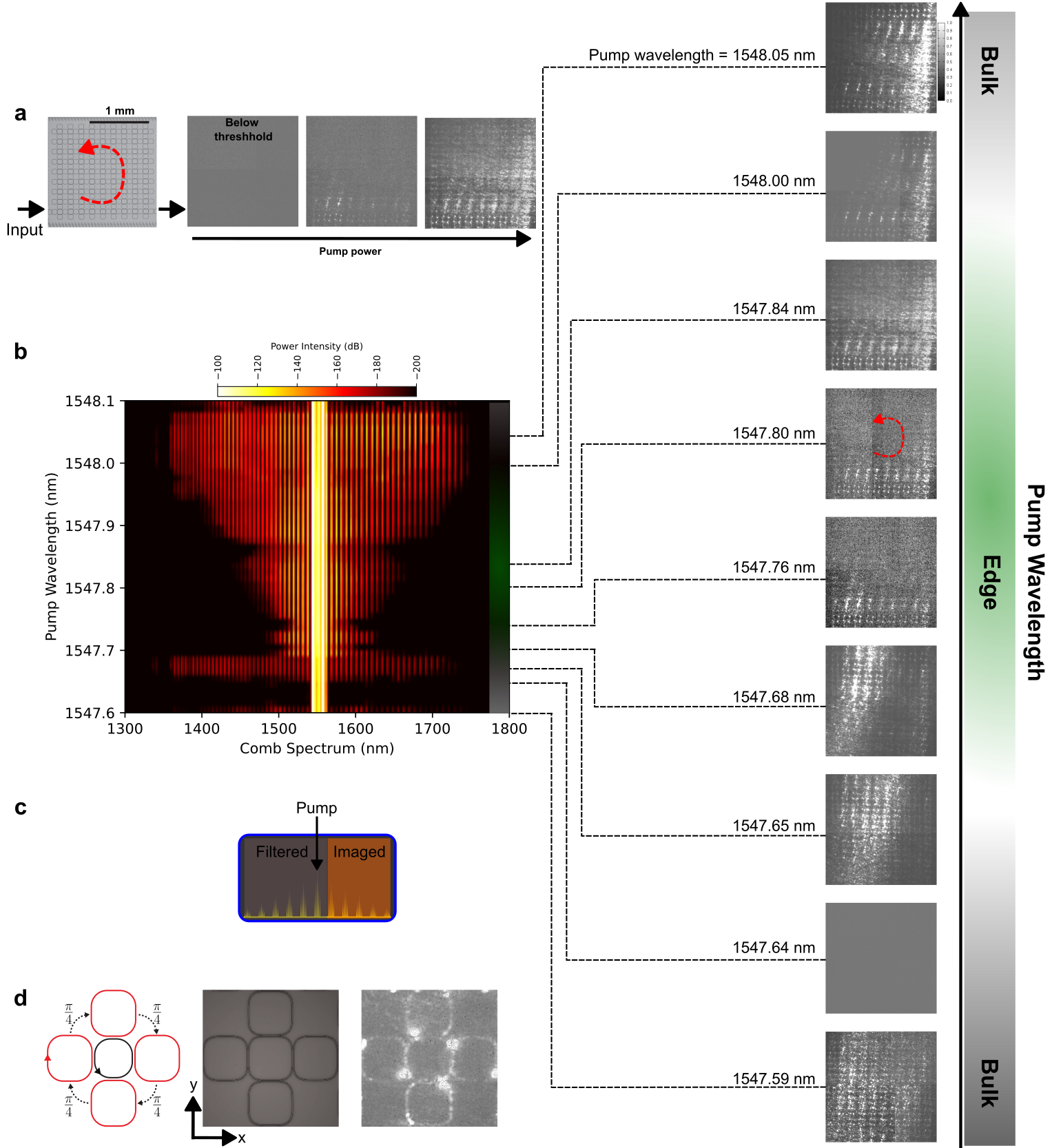}
     \caption{\textbf{Detailed nonlinear imaging.} (a) Pump power dependence of the nonlinear spatial imaging for an edge comb. (b) Measured spatial imaging of several bulk and edge combs. (c) Schmetaic of the integration bandwidth used for the top-down imaging. (d) Nonlinear imaging of a single AQH plaquette.} 
     \label{Fig:Non_Img}
 \end{figure*}

Figure~\ref{Fig:Non_Img} shows the measurement details for the spatial intensity profiles of the generated frequency combs. Figure~\ref{Fig:Non_Img}a shows the pump power dependence of an edge comb (pump wavelength 1547.83 nm). Since the pump is filtered and only part of the generated comb is sent to the IR camera, below the OPO threshold no light is seen. Above the OPO threshold, however, proportional to the pump power, scattered out-of-plane light can be observed.

Figure~\ref{Fig:Non_Img}b shows several intensity profiles for different bulk and edge combs. The strong edge confinement of the edge combs in contrast to the bulk combs can be clearly seen. The same spectral filtering that was applied to all the imaging cases is also shown schematically in Figure~\ref{Fig:Non_Img}c. Moreover, highly magnified imaging of the scattered light from a single plaquette of the AQH is also shown in Figure~\ref{Fig:Non_Img}d.


\newpage
\subsection*{  S8: Nestedness of the combs}

\begin{figure*}[t]
     \centering     \includegraphics[width=0.99\textwidth, angle = -90]{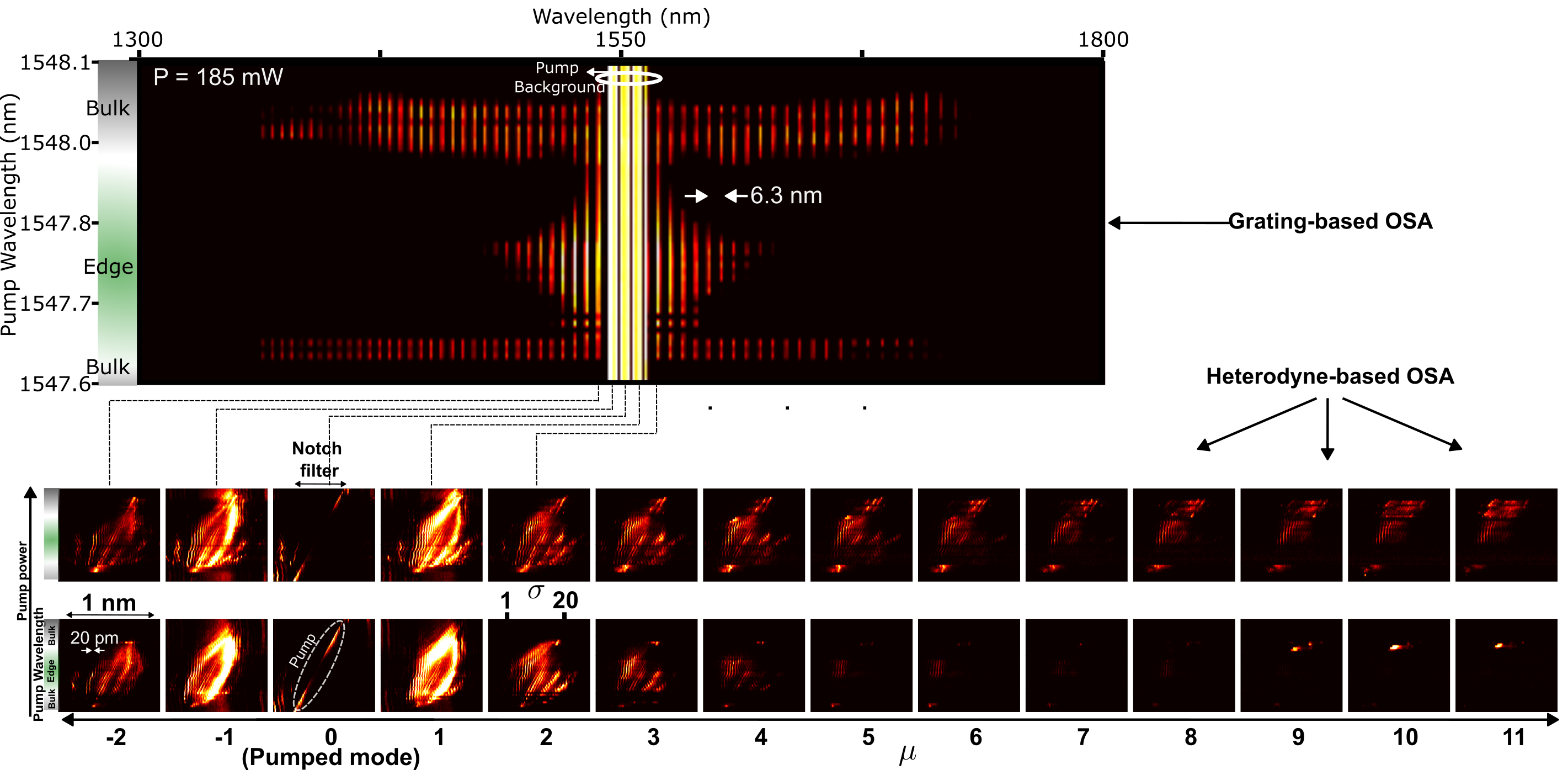}
     \caption{\textbf{Nestedness (nested mode number $\sigma$) as a function of pump power, pump wavelength, and single ring comb tooth number $\mu$. 
     }} 
     \label{Fig:nestedness}
 \end{figure*}

A detailed measurement of the nestedness of the topological combs is presented in Figure~\ref{Fig:nestedness}. Qualitatively, comparable nesting characteristics are observed across all the single ring modes, $\mu$, which are within the bandwidth of our high-resolution OSA (1520 nm to 1630 nm). For edge combs, the comb teeth are nested. As the pump is tuned into the adjacent two bulk bands, the nesting diminishes. Moreover, the data shown for two different pump powers shows no notable change in the nestedness of the combs.


\subsection*{  S9: Optical linewidth Analysis}

\begin{figure*}[t]
     \centering     \includegraphics[width=0.99\textwidth]{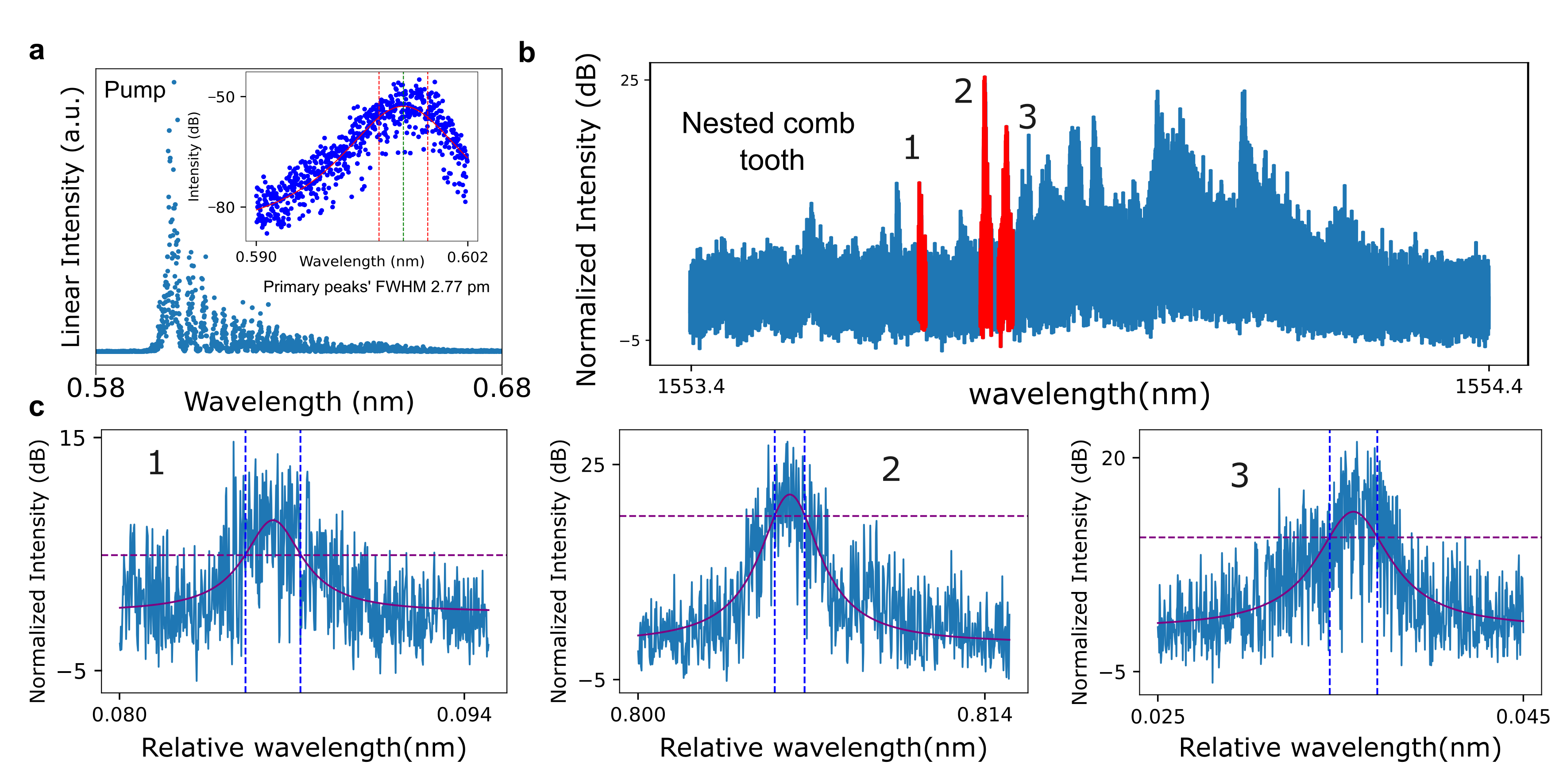}
     \caption{\textbf{Optical linewidth analysis.
     } (a) A typical unfiltered spectrum of the pump laser. The FWHM of the primary peak is 2.77 pm, with an estimated error of $\pm 1.3$ pm from the Lorentzian fit. The inset shows the Lorentzian fit of the primary peak. (b) Typical nested comb teeth, where three nested teeth are highlighted in red for exemplary fit analysis. (c) Lorentzian linewidth fit of the selected comb teeth, as highlighted in red in panel b. It is shown that similar to the linear spectrum, nested teeth on the shorter wavelength side are more resolvable, therefore in this study, we rule out invalid fittings typically at higher wavelength.} 
     \label{Fig:optical_linewidth}
 \end{figure*}

Using the high-resolution (0.4 pm) heterodyne-based OSA, we performed detailed optical linewidth analysis of the pump laser, as well as the nested comb teeth. Figure~\ref{Fig:optical_linewidth}a shows the pump spectrum, which has several peaks, with most of the power concentrated at its dominant primary peak. The Lorentzian fit of the primary peak reveals a FWHM of around 2.77 pm, with an estimated error of $\pm 1.3$ pm. Figure~\ref{Fig:optical_linewidth}b-c shows a typical fitting procedure of the nested comb teeth in our work.  


\newpage
\subsection*{  S10: Pump spectral and temporal calibration}

\begin{figure*}[t]
     \centering     \includegraphics[width=0.99\textwidth]{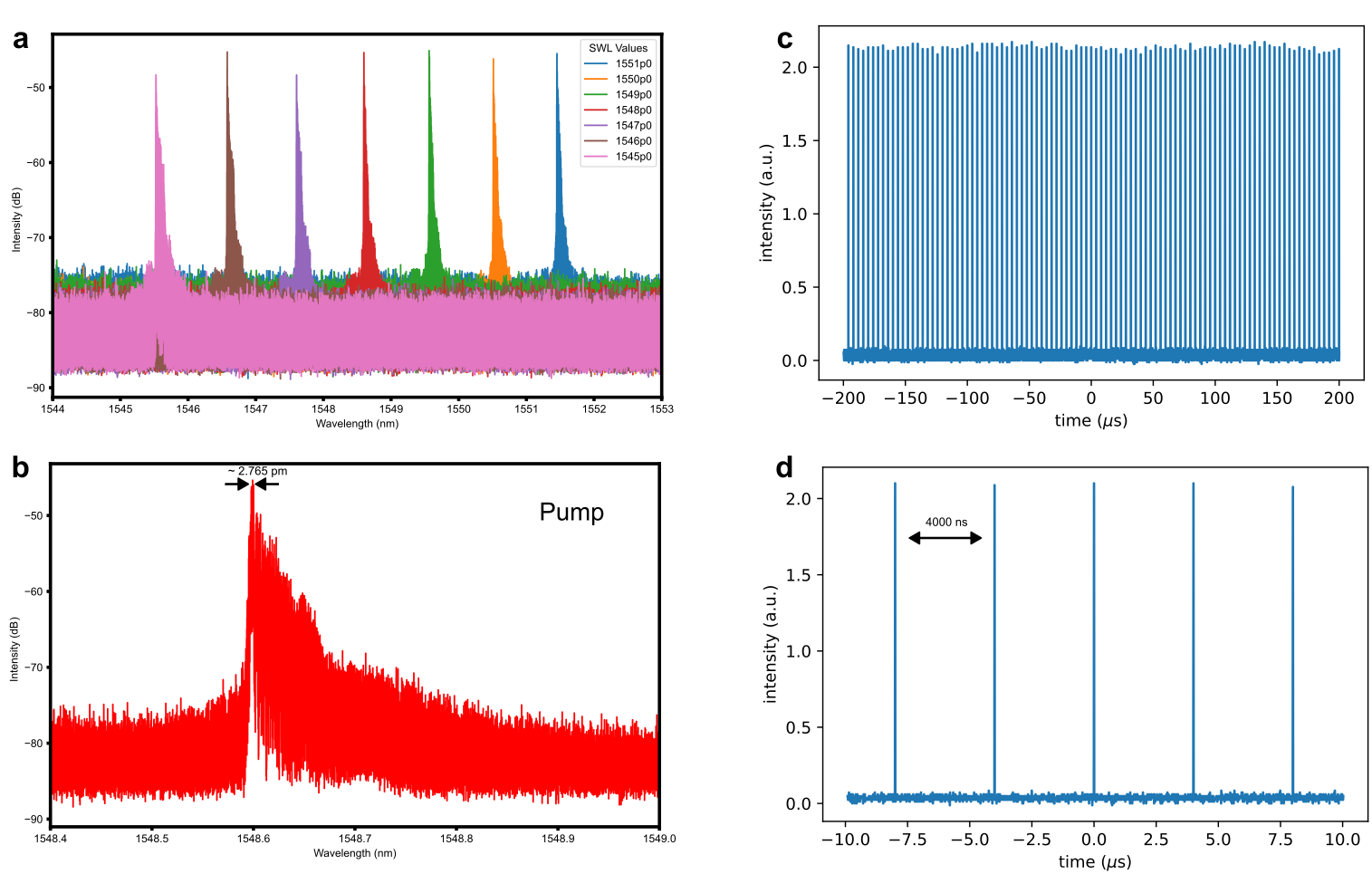}
     \caption{\textbf{Pump Calibration.} (a) Pump spectra at different pump wavelengths over its 6 nm tuning range, and a (b) zoom-in at 1548 nm. Pump temporal pulses, measured with the oscilloscope over many (c) and (d) zoom-in over 5 pulses. Arrows indicate the repetition rate of the pump. } 
     \label{Fig:Pump_rep}
 \end{figure*}

We characterized the pump with the heterodyne-based OSA at several pump wavelengths as shown in Figure~\ref{Fig:Pump_rep}a-b. The pump was also measured directly by sending it to the PD and monitoring the RF output with the oscilloscope, which is shown in Figure~\ref{Fig:Pump_rep}c-d. The 250 kHz (4000 ns) repetition rate of the pump can also be seen.   


%
\subsection*{  S11: Fourier transform}

\begin{figure*}[t]
     \centering     \includegraphics[width=0.99\textwidth]{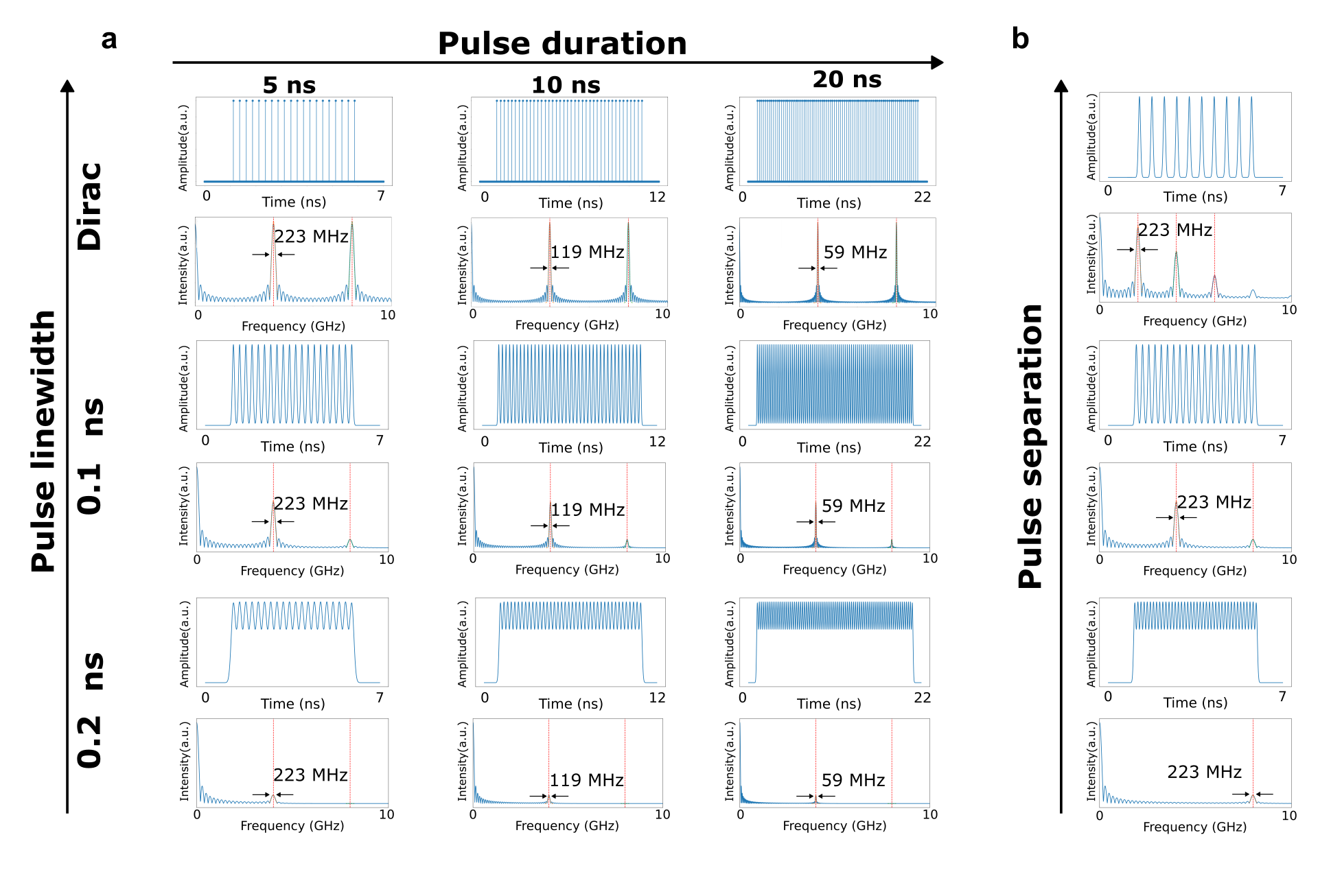}
     \caption{\textbf{The Fourier transform (FT) limit.
     } (a) The FT of the pulse as a function of pulse linewidth (vertical) and pulse duration (horizontal). (b) The FT of the pulse as a function of pulse separation.} 
     \label{Fig:limit}
 \end{figure*}

The finite duration of our pulsed pump puts a lower bound on the frequency resolution in our ESA measurements. This is shown for different scenarios in Figure~\ref{Fig:limit}. With our 5 ns pulse, the round-trip time of around 300 ps to 400 ps, a frequency resolution limit of around 223 MHz is calculated. We note that this limit only depends on the pulse duration, diminishing for longer pulse durations and vice versa. We also note that this limit doesn't depend on the pulse separation as shown in Figure~\ref{Fig:limit}b. 


%
\subsection*{  S12: Notch filter }

\begin{figure*}[t]
     \centering     \includegraphics[width=0.99\textwidth]{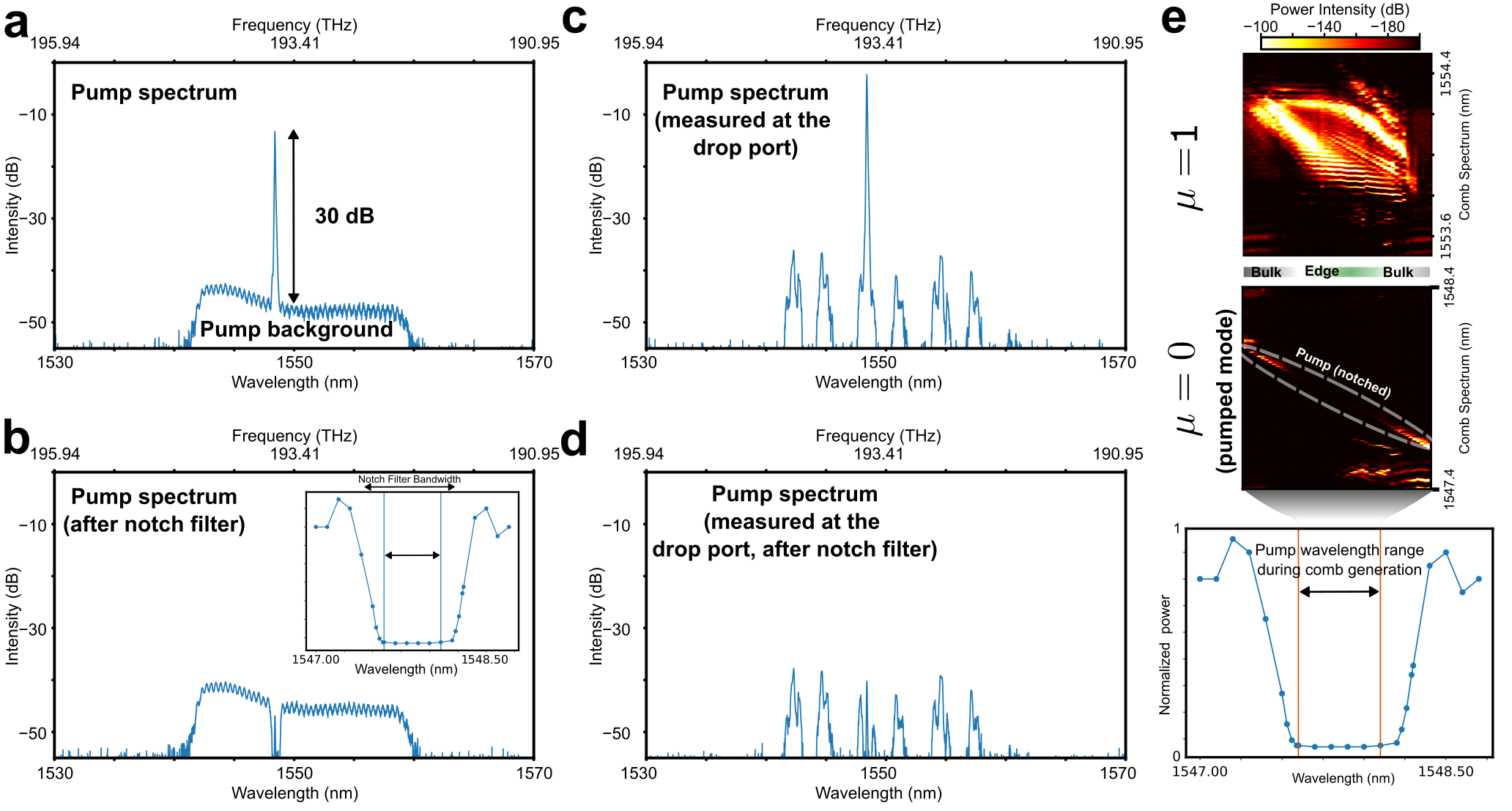}
     \caption{\textbf{Pump characterization and the notch filter.} A typical pump spectrum directly measured with the OSA, (a) without, and (b) with notch filtering. (c-d) The corresponding measurements to (a-b) after the pump is traveled through the edge of the AQH lattice in the linear regime. (e) Bottom: the tunability of the notch filter in the nonlinear regime as a function of pump wavelength over the nonlinear tuning ranges in this work. Middle: the pumped mode after being notched. Top: the first tooth $\mu = 1$ of the comb.} 
     \label{Fig:notch}
 \end{figure*}

In our nonlinear measurement, we filter out the pump optionally with a tunable notch filter and mostly only analyze the remaining part of the comb spectrum. The filter is chosen such that it has a consistent and very high extinction ratio over the full pump wavelength range used in our work, which is 1547.6 nm to 1548.1 nm. The full calibration of the notch filter of the pump with and without going through the device in the linear regime, as well as in the above-OPO regime, is shown in Figure~\ref{Fig:notch}.

\newpage
\subsection*{  S13: OSA, ESA and PD measurements of the combs}
\begin{figure*}[t]
     \centering     \includegraphics[width=0.99\textwidth]{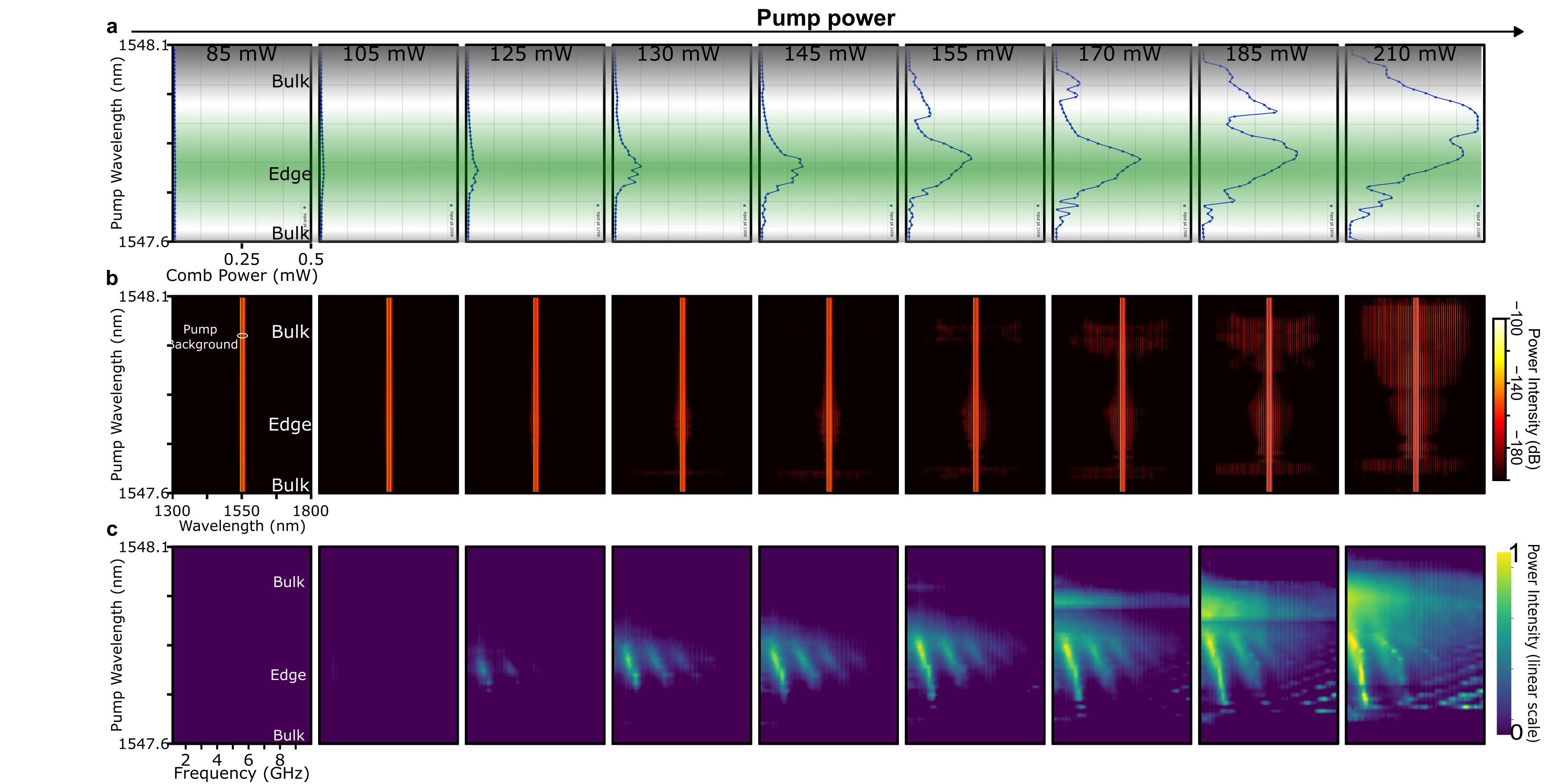}
     \caption{\textbf{OSA, ESA, and PD measurements of the combs.} Measured (a) optical power, (b) OSA, and (c) ESA spectrum of the frequency combs, as a function of pump power (horizontal) and pump wavelength (vertical). The pump is notched out before sending the spectrum to the OSA and a PD monitors the the power directly for optical power measurements. For the ESA measurements, the RF output of the PD was sent to the ESA. } 
     \label{Fig:osa_esa_pd}
 \end{figure*}

The detailed pump power and pump wavelength dependence of the topological combs are summarized in Figure~\ref{Fig:osa_esa_pd}. The top row shows the measured comb powers as a function of pump wavelength over the 1547.6 nm to 1458.1 nm range of the nonlinear measurements. The edge combs have higher power than the bulk counterparts except for the very high pumping regime. The middle and bottom panels show the corresponding OSA and ESA spectrums. The pump was notched out in all of these measurements.


%
\subsection*{  S14: Quadratic fitting of the comb optical linewidth}

\begin{figure*}[t]
     \centering     \includegraphics[width=0.3\textwidth]{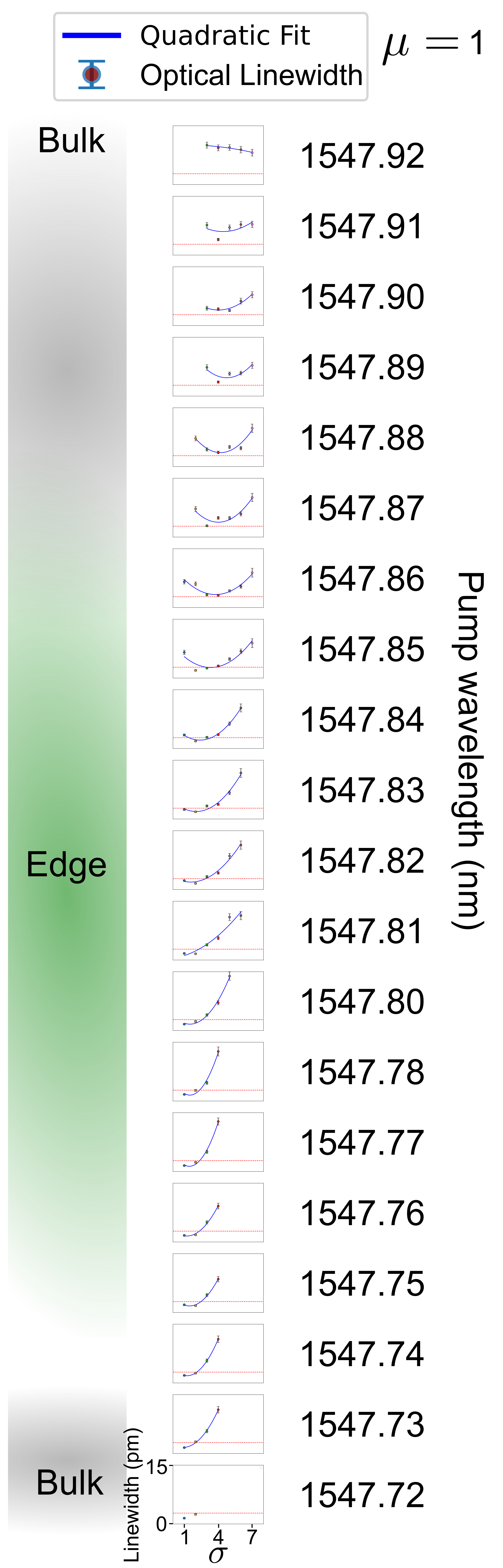}
     \caption{\textbf{Quadratic fitting of the comb optical linewidth.
     } Optical linewidth analysis of the nested teeth $\sigma$ 1,2,3,4,5,6 and 7 of the combs, for a selected longitudinal mode of the single ring $\mu = 1$, for several pump wavelengths. The blue curves and the red dashed lines are the quadratic fit and the pump's linewidth, respectively.}

     \label{Fig:quadratic}
 \end{figure*}

We analyzed the optical linewidth of nested comb lines for a selected single ring mode $\mu = 1$. The results for the nested teeth $\sigma =$ 1 to 7 are shown in Figure~\ref{Fig:quadratic} for several pump wavelengths. We clearly observe a quadratic form (with some variation) of the optical linewidth, which goes close to the pump's linewidth near the pumped mode for most cases. Moreover, the tooth with the minimum linewidth follows the pump wavelength sweeping direction. This behavior is only observed for the edge combs, and the bulk counterparts lack such features.

\begin{figure*}[t]
     \centering     \includegraphics[width=0.99\textwidth]{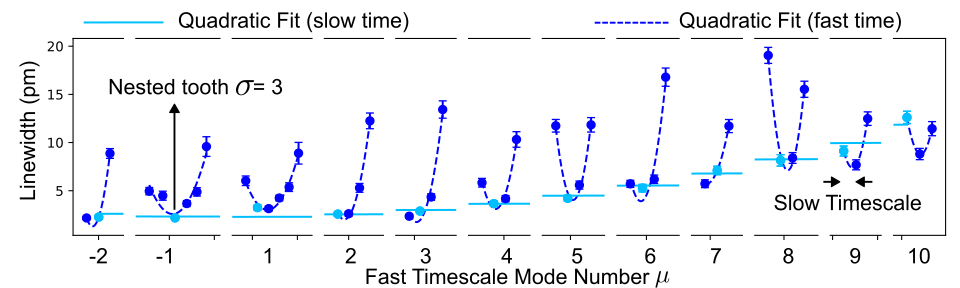}
     \caption{\textbf{Quadratic fitting of the comb optical linewidth over both fast and slow time scales.
     } Optical linewidth analysis of the nested comb teeth $\sigma$ across several longitudinal modes of the single ring $\mu$. The pump wavelength of 1548.6 nm is chosen for this plot, which corresponds to an edge comb. The gaps in the x-axis indicated the zoomed-in windows for each mode. The Lorentzian fitting errors are also shown for each data point.}

     \label{Fig:quadratic2d}
 \end{figure*}

Moreover, Figure~\ref{Fig:quadratic2d} shows the quadratic variation of the comb tooth optical linewidth over both fast and slow timescales for an edge comb. This serves as direct evidence for simultaneous mode-locking over both timescales. 

\begin{figure*}[t]
     \centering     \includegraphics[width=0.99\textwidth]{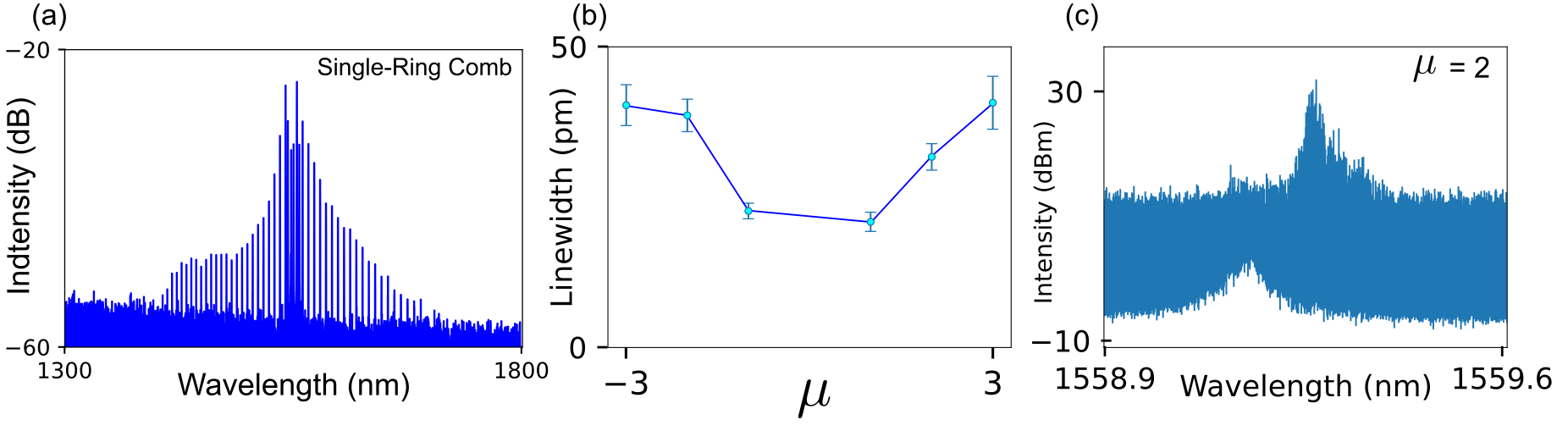}
     \caption{\textbf{Comb optical linewidth analysis for a single-ring race-track device.
     } (a) Generated comb in a single comparable race-track ring resonator. (b) Optical linewidth analysis of the comb tooth across several longitudinal modes of the single ring, which shows much broader linewidth compared to the pump, and lacks a quadratic trend. The Lorentzian fitting errors are also shown. (c) Zoom-in of a typical comb tooth (tooth 2), showing the absence of nestedness.}

     \label{Fig:single}
 \end{figure*}

Further, we compare the linewidth of our topological combs with a comparable (with an identical design) single-ring race-track counterpart. Figure~\ref{Fig:single}a shows a generated comb using the same pump laser in our experimental setup. As shown in Figure~\ref{Fig:single}b, the optical linewidth of the comb teeth lacks the quadratic trend of the topological counterpart and also exhibits a much broader linewidth compared to the pump's linewidth. Moreover, as shown in Figure~\ref{Fig:single}a, nested structure is absent in the comb teeth of the single ring comb. We note that the absence of a quadratic trend across the single-ring comb optical linewidth can be in principle further substantiated by performing the linewidth analysis over a broader range of comb teeth. However, the low SNR of the comb lines away from the pump limits our ability to perform reliable fitting analysis.


\newpage
\subsection*{  S15: Temporal analysis of the pump power dependence of the bulk combs}

\begin{figure*}[t]
     \centering     \includegraphics[width=0.9\textwidth]{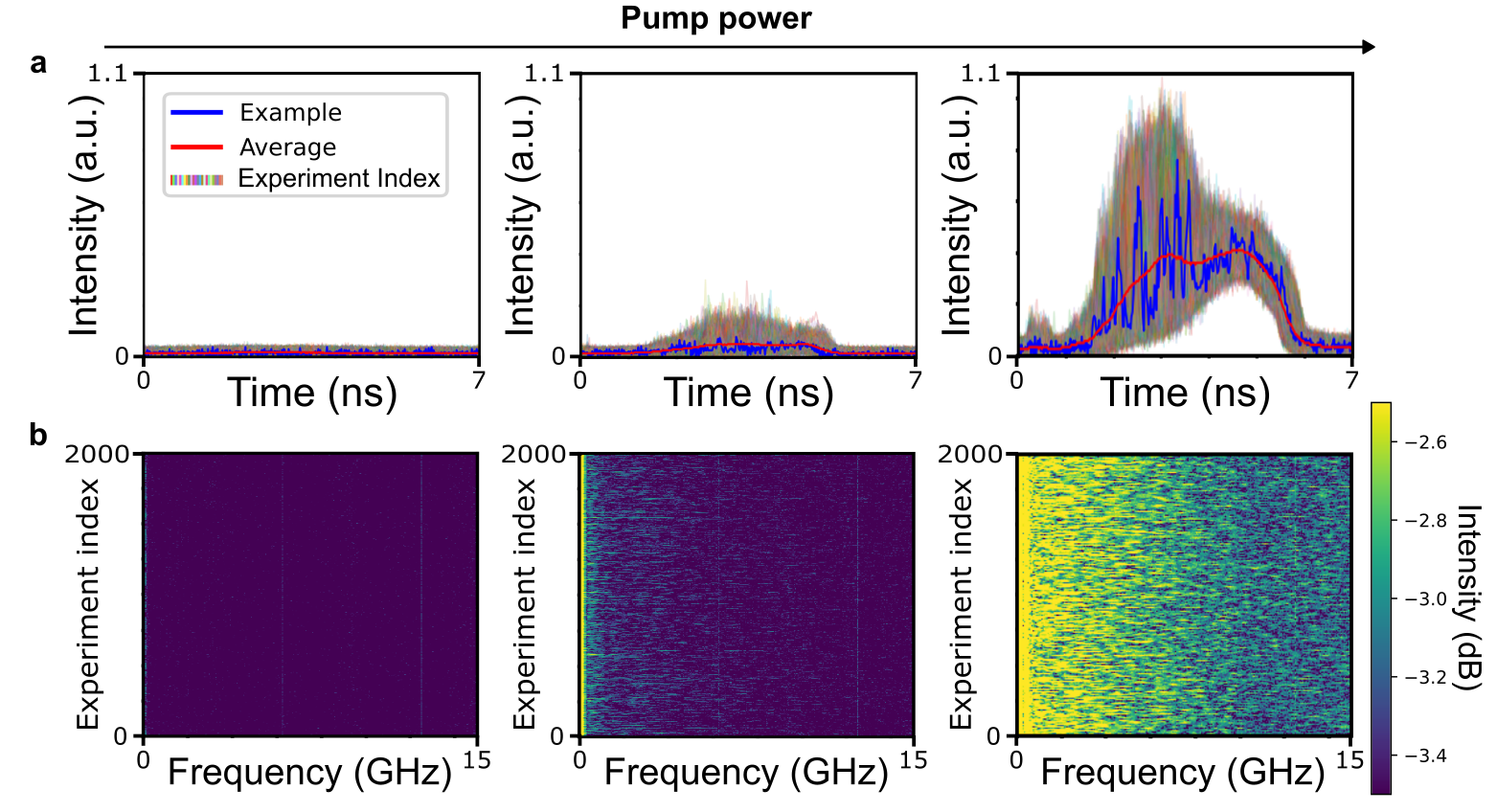}
     \caption{\textbf{Temporal analysis of the pump power dependence of the bulk combs.
     } (a) Oscilloscope measurements of a bulk comb as a function of pump power. The plots include 2000 repeats of the experiment, with one example shown in blue and the average shown in red. (b) The corresponding Fourier transform of the data in (a). The pump power from left to right: 125 mW, 155 mW, 210 mW} 
     \label{Fig:scope_bulk_power}
 \end{figure*}

For a typical bulk comb, the temporal dynamic of the comb is measured as a function of pump power. The results are shown in Figure~\ref{Fig:scope_bulk_power}. It can be seen that no appreciable temporal patterns emerge above the OPO threshold, leading to no appearance of a well-defined frequency component in the (Fourier transform) FT analysis.


%
\subsection*{  S16: Temporal analysis of the pump power dependence of the edge combs}

\begin{figure*}[t]
     \centering     \includegraphics[width=0.99\textwidth]{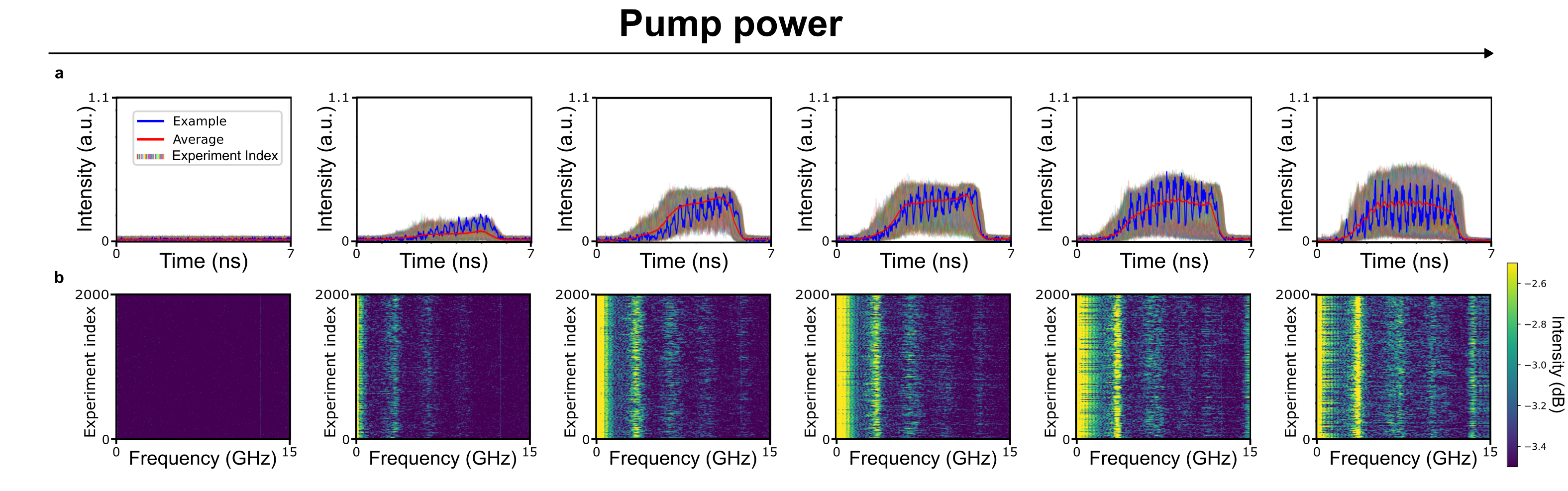}
     \caption{\textbf{Temporal analysis of the pump power dependence of the edge combs.
     } (a) Oscilloscope measurements of an edge comb as a function of pump power. The plots include 2000 repeats of the experiment, with one example shown in blue and the average shown in red. (b) The corresponding Fourier transform of the data in (a). The pump power from left to right: 85 mW, 125 mW, 155 mW, 170 mW, 185 mW, 210 mW} 
     \label{Fig:scope_edge_powe}
 \end{figure*}

For a typical edge comb, the temporal dynamics of the comb are measured as a function of pump power. The results are shown in Figure~\ref{Fig:scope_edge_powe}. Unlike the bulk combs, it can be seen that temporal patterns emerge above the OPO threshold, leading to the appearance of well-defined frequency components in the FT analysis.
 

\subsection*{  S17: Temporal analysis of the pump power and pump wavelength dependence of the edge and bulk combs}

\begin{figure*}[t]
     \centering     \includegraphics[width=0.99\textwidth]{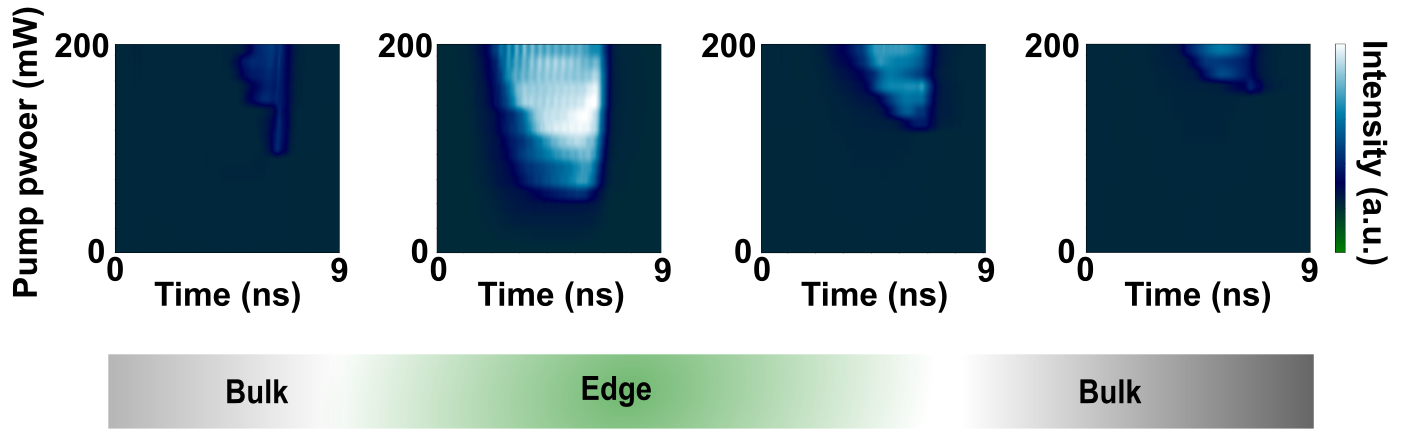}
     \caption{\textbf{Temporal analysis of the pump power and pump wavelength dependence of the edge and bulk combs.
     } Oscilloscope measurement of the combs as a function of pump power and pump wavelength. Above the OPO threshold (80 mW), temporal patterns form only for the edge combs. The pump wavelength from left to right: 1547.65 nm, 1547.76 nm, 1548.02 nm, 1548.05 nm} 
     \label{Fig:scope_edge_bulk_powe}
 \end{figure*}

For several edge and bulk combs, the temporal dynamics of the comb are measured as a function of pump power. The results are shown in Figure~\ref{Fig:scope_edge_bulk_powe}. The sharp contrast between edge combs and their bulk counterparts can be seen.


\subsection*{  S18: Temporal analysis of the combs as a function of pump wavelength.}

\begin{figure*}[t]
     \centering     \includegraphics[width=0.99\textwidth, angle = -90]{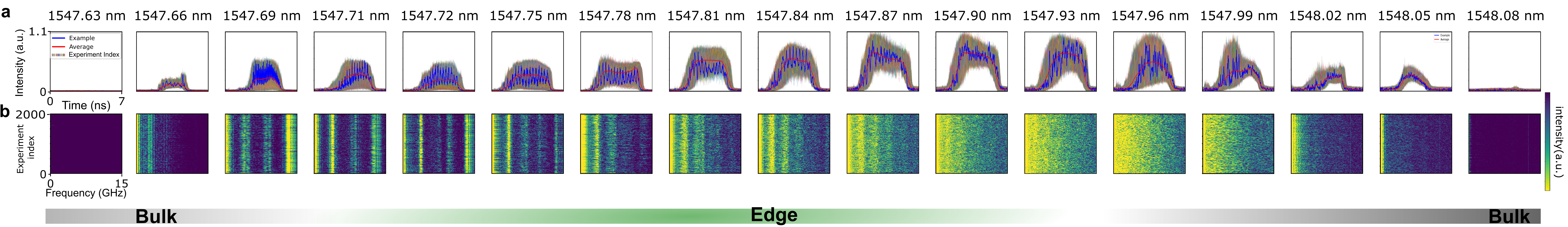}
     \caption{\textbf{Temporal analysis of the combs as a function of pump wavelength.
     } (a) Oscilloscope measurements of an edge comb as a function of pump wavelength for a fixed pump power of 185 mW. The plots include 2000 repeats of the experiment, with one example shown in blue and the average shown in red. (b) The corresponding Fourier transform of the data in (a).} 
     \label{Fig:scope_edge_detuning}
 \end{figure*}

For a fixed pump power of 185 mW, the temporal dynamics of the combs are measured as a function of pump wavelength. The results are shown in Figure~\ref{Fig:scope_edge_detuning}. The gradual emergence of temporal patterns as the pump is tuned towards the edge band can be seen.

\subsection*{  S19: Transient dynamics of the lattice}
\begin{figure*}[t]
     \centering     \includegraphics[width=0.99\textwidth, angle = 0]{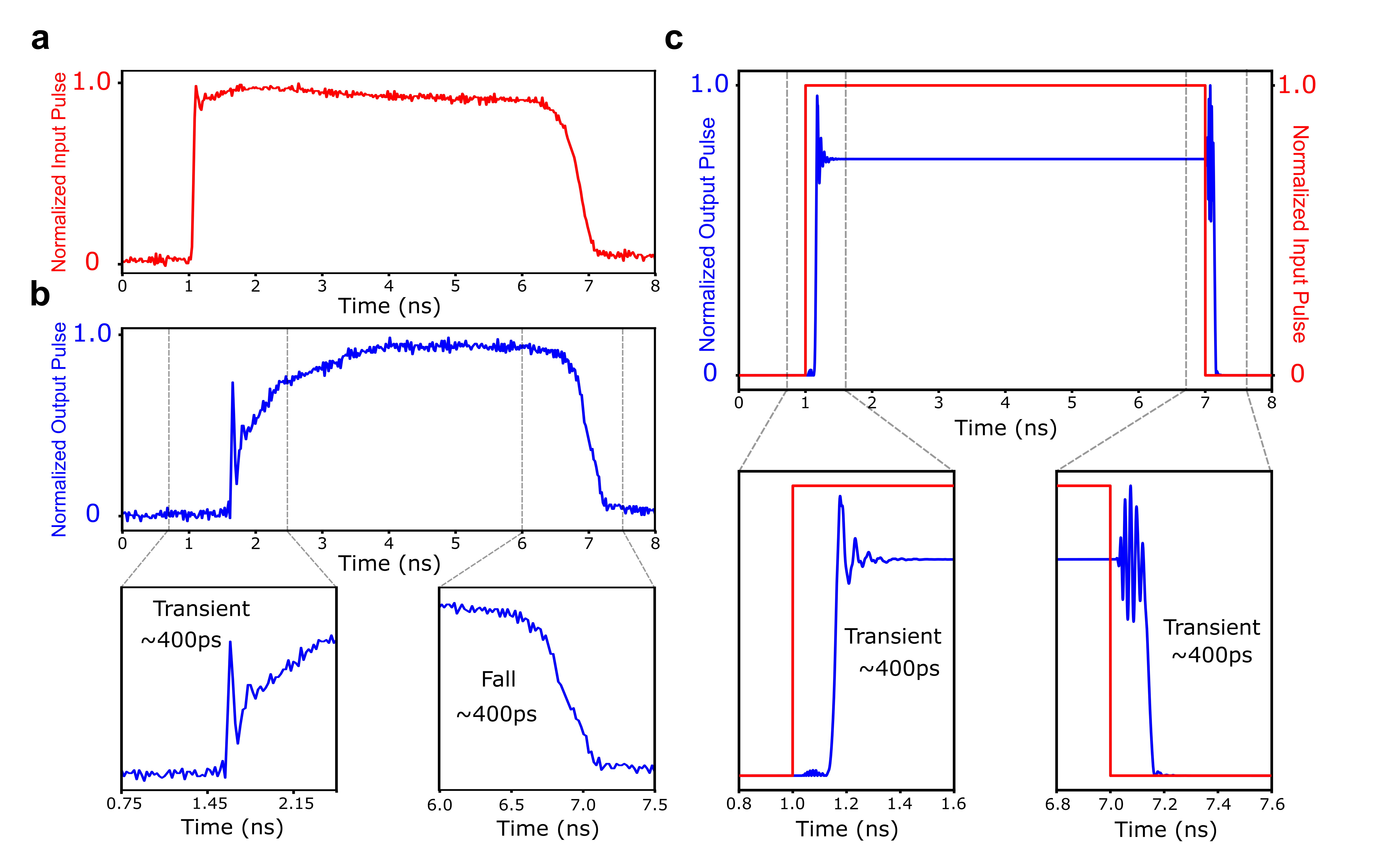}
     \caption{\textbf{Transient dynamics of the lattice
     .} (a) Direct oscilloscope measurement of the laser pulse without sending into the chip. (b) Oscilloscope measurement of the laser pulse after the chip, collected from the drop port. The power of the pulse is below threshold. Note that panels (a) and (b) are two independent measurements with unmatched x-axis. (c) The corresponding simulation of transient behavior with input in red and output in blue.} 
     \label{Fig:transient}
\end{figure*}

Due to the pulsed nature of our laser, one important question is whether the transient behavior, namely the response time of the system until it reaches steady state, is comparable to the 5 ns pulse duration. Figure~\ref{Fig:transient} shows the transient dynamics of our system from both experiment and simulation. It can be seen that from both the experiment and simulation, the duration of transient behavior that is dominated by damped oscillations is about 400 ps, which is at least an order of magnitude lower than the 5 ns pulse duration. The physical reason is that both the intrinsic and extrinsic losses of individual rings are high: the loaded quality factor for a single ring is measured to be only a few thousand.

\subsection*{  S20: Filtered-comb experiment}

The signatures of mode locking have been demonstrated from both time and frequency domain measurements in the main text. However, several limitations are posed by our laser. Firstly, the finite pulse duration sets a lower limit for frequency-domain resolution, causing our ESA beat notes to be broader than 223 MHz. Secondly, the broad laser background in the frequency domain diminishes the quality of imaging and adds substantial noise in temporal measurements. To address these issues, we split the comb with 1600 nm long pass and short pass filters, and perform the measurements as described in the main text on each part of the comb. Additionally, we use another device with the same design parameters to show the reproducibility of the results demonstrated in the main text.

\begin{figure*}[t]
     \centering     \includegraphics[width=0.99\textwidth, angle = 0]{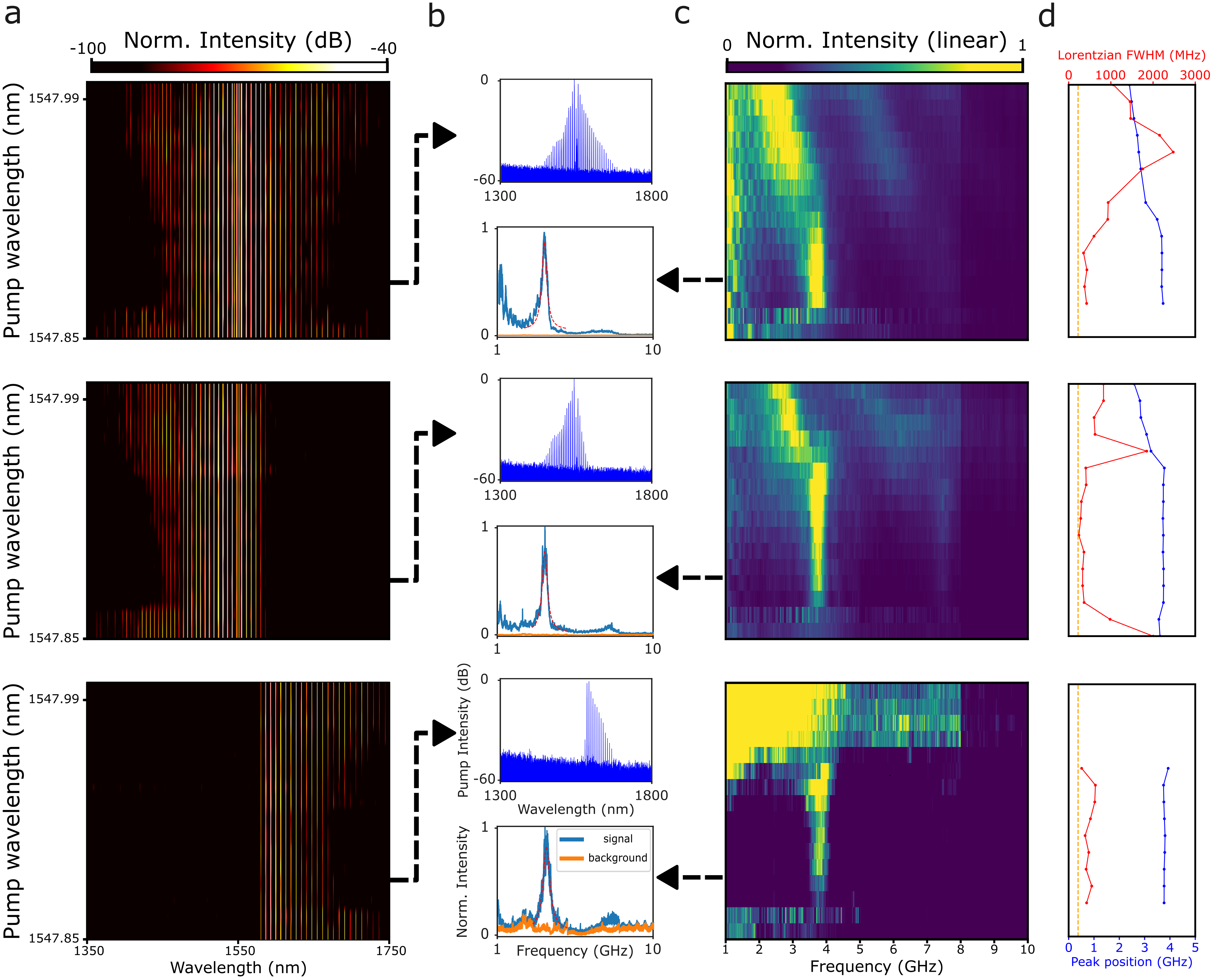}
     \caption{\textbf{Filtered measurements for OSA and ESA
    } (a,c) Comb spectrum (OSA) and frequency spectrum (ESA) as a function of pump wavelength (b) OSA and ESA measurement for a fixed pump wavelength. From top to bottom, the measurements are performed on unfiltered, 1600 nm short pass filtered and 1600 nm long pass filtered combs.} 
     \label{Fig:filtered_OSA_ESA}
 \end{figure*}

Figure~\ref{Fig:filtered_OSA_ESA} shows the OSA and ESA measurements for combs with and without filters. The OSA measurements clearly demonstrate the effect of the filters we use, where the pump background is included in the 1600 nm short pass spectrum, but completely filtered out for the 1600 nm long pass spectrum. For all filtering schemes, we see the same ESA beat notes around 4 GHz with near-transform-limit linewidth, corresponding to the slow-time repetition rate.

\begin{figure*}[t]
     \centering     \includegraphics[width=0.99\textwidth, angle = 0]{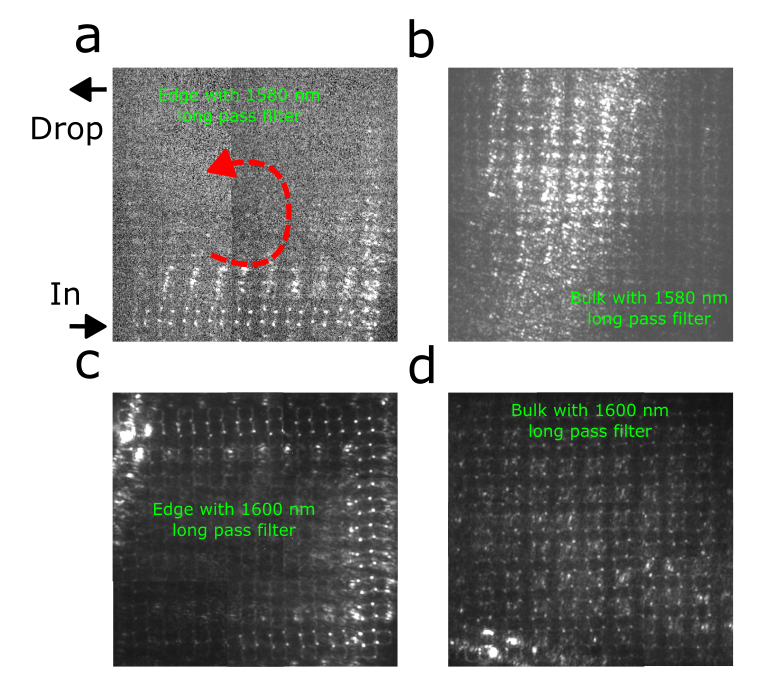}
     \caption{\textbf{Imaging of 1600 nm long pass filtered combs.} (a) Image of edge states for combs with the old 1580 nm long pass filter. (b) Image of bulk states for combs with the old 1580 nm long pass filter. (c) Improved edge state imaging with a new 1600 nm filter on a new device. (d) Improved bulk state imaging with new 1600 nm filter on a new device.} 
     \label{Fig:filtered_imaging}
\end{figure*}

Next, we perform imaging of the system with the 1580 nm and 1600 nm long pass filter. The pump background is completely filtered out with the 1600 nm filter, but remains for 1580 nm filter. The results are shown in Figure~\ref{Fig:filtered_imaging}. We notice a considerable improvement in the image when we filter out the complete pump background.

We then generate combs with a fixed wavelength in the edge band, and proceed to analyze both filtered parts of the comb in the time domain with the fast (20 GHz) oscilloscope, as shown in Figure~\ref{Fig:filtered_Osc_analysis}. By removing the pump background with a 1600 nm long pass filter, and contrasting the result against the 1600 nm short pass filter, where the pump background remains, we have greatly improved the signal to noise ratio of the measurement, as shown in panel a. Remarkably, by fitting the temporal pulses, we show that these pulses can go as narrow as $\approx$ 50 ps, which is a limit set by the oscilloscope (20 GHz). To confirm that these pulses do not come from transient behavior, we fix the wavelength and pump the system below OPO threshold. The measured temporal dynamics in panel b indeed shows that the comb pulses do not come from transient behavior.

\begin{figure*}[t]
     \centering     \includegraphics[width=0.99\textwidth, angle = 0]{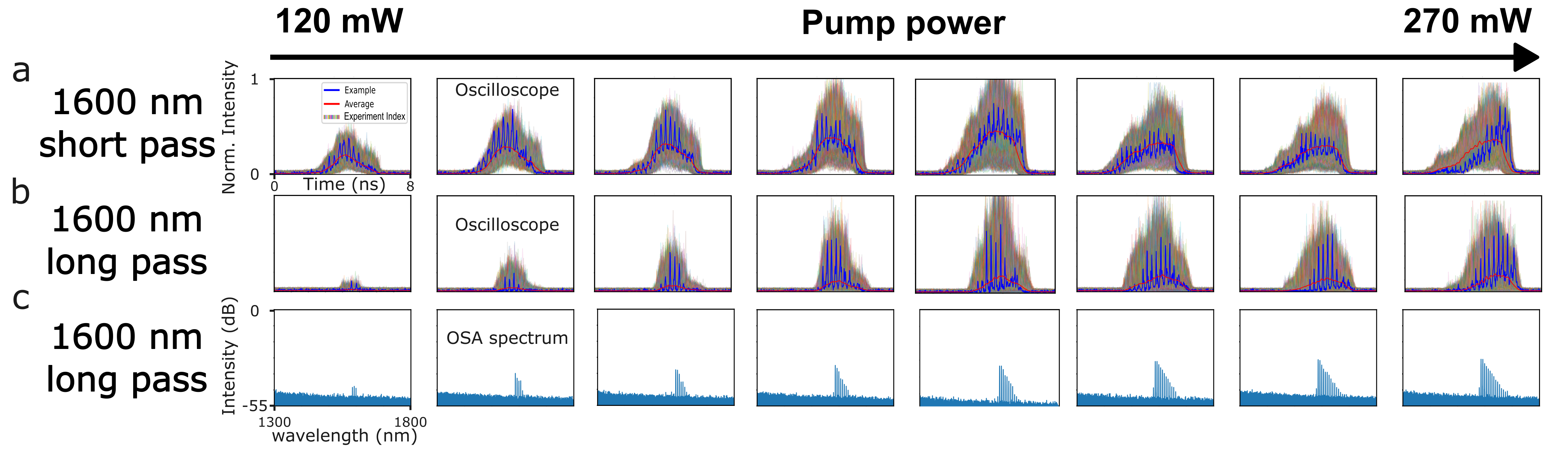}
     \caption{\textbf{Temporal measurement of filtered combs as a function of pump power 
     } For a fixed wavelength in edge state, the pump power is swept from 120 mW to 270 mW. (a) Oscilloscope measurements for 1600 nm short pass filtered comb. (b) Oscilloscope measurements for 1600 nm long pass filtered comb. (c) The corresponding filtered comb spectrum for panel (b)} 
     \label{Fig:filtered_Osc_power_sweep}
\end{figure*}

\begin{figure*}[t]
     \centering     \includegraphics[width=0.99\textwidth, angle = 0]{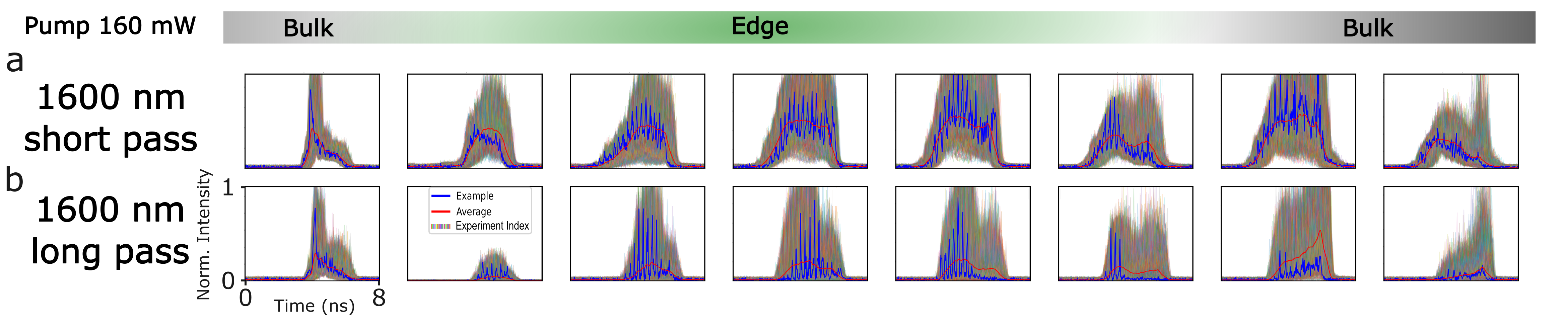}
     \caption{\textbf{Temporal measurement of filtered combs as a function of wavelength 
     } For a fixed pump power, the pump wavelength is swept from one bulk band, across the edge band, and then to the other bulk band. (a) Oscilloscope measurements for 1600 nm short pass filtered comb. (b) Oscilloscope measurements for 1600 nm long pass filtered comb. } 
     \label{Fig:filtered_Osc_wavelength_sweep}
\end{figure*}

Similar to Sections S16 and S17, we have performed sweeps on both pump power (Figure~\ref{Fig:filtered_Osc_power_sweep}) and pump wavelength (Figure~\ref{Fig:filtered_Osc_wavelength_sweep}), showing again the sharp contrast between bulk combs and edge combs, as well as the power dependence of the edge combs in time domain.  

To simultaneously show the temporal signatures of mode locking for both parts of the comb (above and below 1600 nm), we have added a supplementary video 1. The video sweeps over 200 independent experiments, showing not only the correlation between two parts of comb in time, but also the repeatability of the results.

\subsection*{  S21: Spacing between oscillating edge modes}
\begin{figure*}[t]
     \centering     \includegraphics[width=0.99\textwidth, angle = 0]{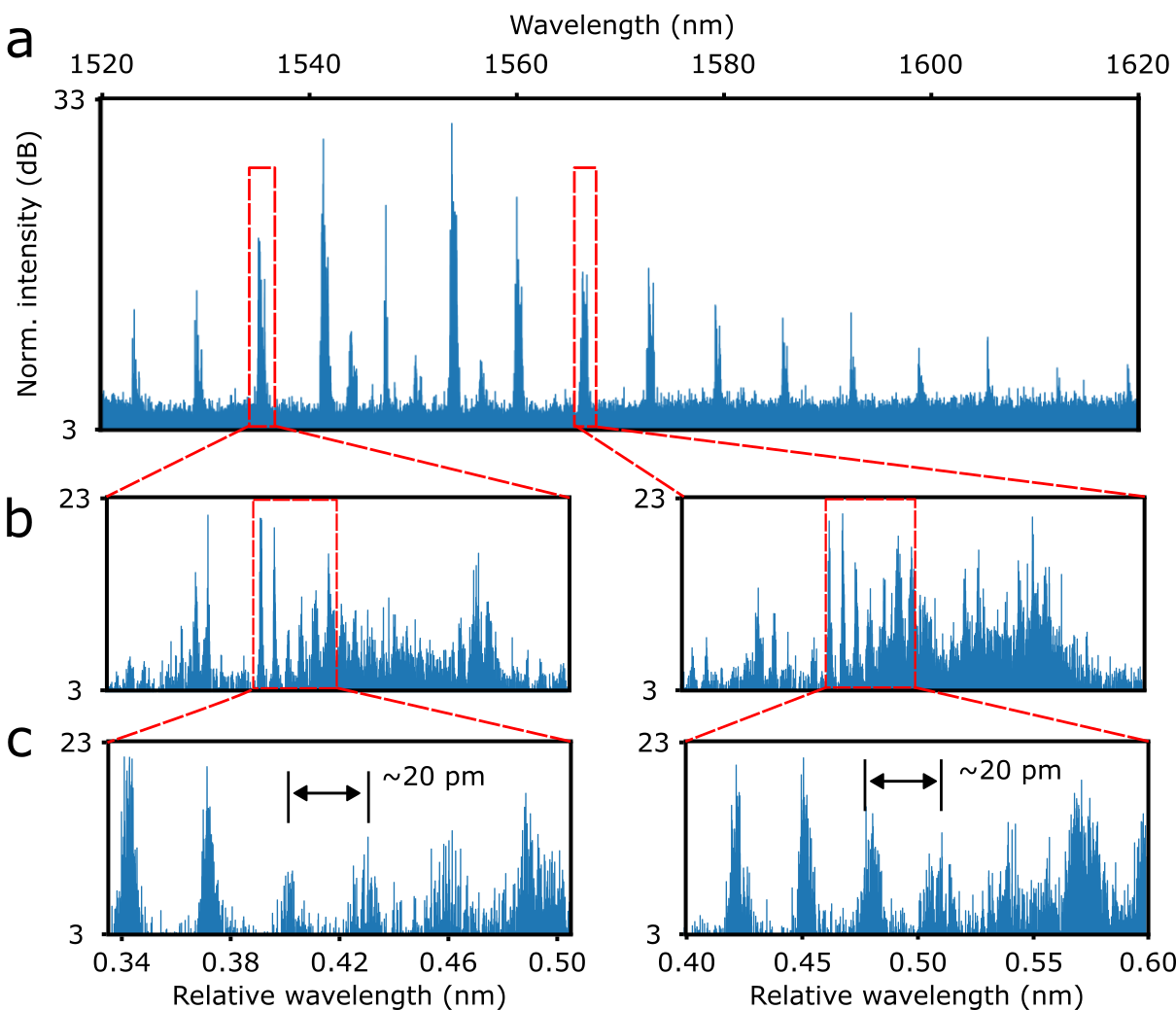}
     \caption{\textbf{Zoomed in demonstration of the slow time-scale.
     }High-Resolution Spectra of Individual Comb Teeth. (a) High resolution OSA spectrum for pump wavelength 1547.78nm. (b) The selected fast timescale modes -2 and 3, zoomed in from (a). (c) The selected slow time-scale modes, zoomed in from (b).  } 
     \label{Fig:20pm}
\end{figure*}

To better demonstrate the spacing between oscillating edge modes demonstrated in Figure~\ref{Fig:osa}d, which is approximately 20 pm, we plot the optical spectrum when we pump in the edge with wavelength 1547.78 nm, as shown in Figure~\ref{Fig:20pm}. Here, we choose two single-ring modes $\mu = -2,3$, and clearly shows that the spacing between super-ring modes $\sigma$ is about 20 pm.

We note that, due to the ultra-high resolution of the OSA (0.04 pm), the transmission spectrum has a very high sampling rate even within a narrow 0.2 nm window. Moreover, since the damage threshold of the grating-based OSA is much lower than conventional grating-based counterparts, we attenuated the input optical signal prior to the OSA, which leads to a low signal-to-noise ratio. These two factors combined give the comb teeth spectrum a slightly noisy and spiky appearance.  

\subsection*{  S22: Limitations of the experiment}
In this study, we identify strong signatures of mode locking with various independent measurements. However, direct phase measurements to show mode locking are not presented, and this is due to several experimental challenges and limitations in the current stage. A detailed breakdown of the challenges are as follows:

\begin{itemize}
    \item \textbf{Coupling 261 rings together.} The topological ring lattice contains hundreds of single rings leading to huge fabrication variations. One has to intentionally design a very low quality factor for individual rings (a few thousand) so that the whole device works.
    \item \textbf{High OPO threshold.} A very low quality factor leads to a very high OPO threshold. In our experiment, the peak power of the pulse has to reach about 100W on threshold. On the other hand, using a CW laser with 100W introduces detrimental heat effects that introduce instability and damage chips. Consequently, we use a pulsed laser with a strong peak power, but manageable average power.
    \item \textbf{Limited pulse duration.} In the experiment, the laser duration is 5 ns, which is more than three orders of magnitude longer than single-ring round trip time, at least an order of magnitude longer than both the super-ring round-trip time and the transient time of the system. Even though such a quasi-CW laser is enough to reveal beat notes on frequency measurements, its limited duration sets a lower resolution in frequency domain, as shown in our ESA measurement.
    \item \textbf{No CW laser.}
    The pulsed nature of the pump stops us from continuously tuning the pump wavelength, which is a common way to find solitons, and is a proposed scheme to find topological solitons in~\cite{mittal2021topological}. Note that we are not working in the same parameter regime as that described in~\cite{mittal2021topological}, therefore we are not claiming the existence of solitons based on this study.
    \item \textbf{Laser jitter.} The laser used in this study jitters in time, as shown in Figure~\ref{Fig:scope}a, where 2000 independent measurements are plotted on top of each other. The jitter of the laser adds to the broadening of the ESA beat note.
    \item \textbf{Laser background.} The laser used in this study has a spectral background that introduces scattering in imaging and adds noise to both frequency and time-domain measurements. To address this issue, we use spectral filters to eliminate the effect of backgrounds.   
\end{itemize}

To provide stronger evidence of mode locking and its type (Turing rolls, topological solitons), much improvement is needed for the experiment. We note that going forward, an important direction is to improve the Q factor of the system to enable narrow-linewidth CW excitation while reducing the lattice size to make sure that all individual rings are properly coupled, which can make the metrology-type applications, as well as the generation of topological solitons more accessible.


\begin{figure*}[t]
     \centering     \includegraphics[width=0.99\textwidth, angle = 0]{fig_SI/FigSI_filtered_Osc_analysis.png}
     \caption{\textbf{Temporal analysis for filtered combs
     } (a) Comparison of temporal measurements for 1600 nm short pass filtered and long pass filtered combs. The inset is the fit for one of the pulses for 1600 nm long pass. (b) Comparison of temporal measurements between below threshold pumping and nonlinearly generated comb signals above 1600 nm, each normalized to themselves.} 
     \label{Fig:filtered_Osc_analysis}
 \end{figure*}

\clearpage
\thispagestyle{empty}

\noindent\textbf{Movie S1. Simultaneous temporal measurement of different parts of the comb.} 
The attached movie S1 shows 200 independent temporal measurements of the comb with the fast (20~GHz) oscilloscope as we pump the edge band and see signatures of mode locking. On each snapshot, we simultaneously measure two parts of topological edge combs using a 1580~nm spectral filter, where the 1580~nm short pass contains parts of the laser background and the 1580~nm long pass contains only the generated comb teeth. It can be seen from the movie that the two parts of the comb are synchronized in time, both giving temporal pulses at the same moment.

\vspace*{\fill}



\end{document}